\documentclass[aps,prx,reprint,superscriptaddress]{revtex4-1}

\usepackage{graphicx}
\usepackage{bm}

\begin{document}


\title{Proximity-induced hidden order transition in a correlated heterostructure Sr$_2$VO$_3$FeAs}

\author{Sunghun Kim}
\thanks{These authors contributed equally to this work}
\affiliation{Department of Physics, Korea Advanced Institute of Science and Technology, Daejeon 34141, Republic of Korea}

\author{Jong Mok Ok}
\thanks{These authors contributed equally to this work}
\affiliation{Center for Artificial Low Dimensional Electronic Systems, Institute for Basic Science, Pohang 37673, Republic of Korea}
\affiliation{Department of Physics, Pohang University of Science and Technology, Pohang 37673, Republic of Korea}

\author{Hanbit Oh}
\affiliation{Department of Physics, Korea Advanced Institute of Science and Technology, Daejeon 34141, Republic of Korea}

\author{Chang-il Kwon}
\affiliation{Center for Artificial Low Dimensional Electronic Systems, Institute for Basic Science, Pohang 37673, Republic of Korea}
\affiliation{Department of Physics, Pohang University of Science and Technology, Pohang 37673, Republic of Korea}

\author{Y. Zhang}
\affiliation{National Laboratory of Solid State Microstructures, School of Physics, Collaborative Innovation Center of Advanced Microstructures, Nanjing University, Nanjing 210093, China}
\affiliation{Advanced Light Source, Lawrence Berkeley National Laboratory, Berkeley, CA 94720, USA}

\author{J. D. Denlinger}
\affiliation{Advanced Light Source, Lawrence Berkeley National Laboratory, Berkeley, CA 94720, USA}

\author{S.-K. Mo}
\affiliation{Advanced Light Source, Lawrence Berkeley National Laboratory, Berkeley, CA 94720, USA}

\author{F. Wolff-Fabris}
\affiliation{Dresden High Magnetic Field Laboratory, Helmholtz-Zentrum Dresden-Rossendorf, Dresden, D-01314, Germany}

\author{E. Kampert}
\affiliation{Dresden High Magnetic Field Laboratory, Helmholtz-Zentrum Dresden-Rossendorf, Dresden, D-01314, Germany}

\author{Eun-Gook Moon}
\email[]{egmoon@kaist.ac.kr}
\affiliation{Department of Physics, Korea Advanced Institute of Science and Technology, Daejeon 34141, Republic of Korea}

\author{C. Kim}
\email[]{changyoung@snu.ac.kr}
\affiliation{Center for Correlated Electron Systems, Institute for Basic Science, Seoul 08826, Republic of Korea}
\affiliation{Department of Physics and Astronomy, Seoul National University, Seoul 08826, Republic of Korea}

\author{Jun Sung Kim}
\email[]{js.kim@postech.ac.kr}
\affiliation{Center for Artificial Low Dimensional Electronic Systems, Institute for Basic Science, Pohang 37673, Republic of Korea}
\affiliation{Department of Physics, Pohang University of Science and Technology, Pohang 37673, Republic of Korea}

\author{Y. K. Kim}
\email[]{yeongkwan@kaist.ac.kr}
\affiliation{Department of Physics, Korea Advanced Institute of Science and Technology, Daejeon 34141, Republic of Korea}
\affiliation{Graduate School of Nanoscience and Technology, Korea Advanced Institute of Science and Technology, Daejeon 34141, Republic of Korea}


\date{\today}

\begin{abstract}
Symmetry is one of the most significant concepts in physics, and its importance has been largely manifested in phase transitions by its spontaneous breaking. In strongly correlated systems, however, mysterious and enigmatic phase transitions, inapplicable of the symmetry description, have been discovered and often dubbed hidden order transitions, as found in, \textit{e.g.}, high-$T_C$ cuprates, heavy fermion superconductors, and quantum spin liquid candidates. Here, we report a new type of hidden order transition in a correlated heterostructure Sr$_2$VO$_3$FeAs, whose origin is attributed to an unusually enhanced Kondo-type proximity coupling between localized spins of V and itinerant electrons of FeAs. Most notably, a fully isotropic gap opening, identified by angle-resolved photoemission spectroscopy, occurs selectively in one of the Fermi surfaces below $T_{\rm HO}$ $\sim$ 150 K, associated with a singular behavior of the specific heat and a strong enhancement on the anisotropic magnetoresistance. These observations are incompatible with the prevalent broken-symmetry-driven scenarios of electronic gap opening and highlight a critical role of proximity coupling. Our findings demonstrate that correlated heterostructures offer a novel platform for design and engineering of exotic hidden order phases.
\end{abstract}


\maketitle


\section{Introduction}

Correlated electron systems often exhibit a variety of self-organized forms with broken symmetry, due to various complex interactions. Usually, the resulting electronic phases have broken time-reversal or lattice symmetry, but sometimes a so-called ``hidden order" phase is stabilized, with no associated broken symmetry identified by the microscopic probes. 
Since the first discovery of the hidden order phase in the heavy fermion superconductor URu$_{2}$Si$_{2}$\cite{Palstra1985,Maple1986a}, tremendous efforts have been made to characterize the intriguing properties of the hidden order phase\cite{Mydosh2011a,Mydosh2020}, some of which are shared with the pseudo-gap phase in underdoped high-$T_C$ cuprates\cite{VojtaRev2009,Sato2017}, the odd-parity phase in doped iridates\cite{Zhao2016}, and Kitaev quantum spin liquid phase in $\alpha$-RuCl$_3$\cite{Banerjee2016}. While a complete understanding of the hidden order phases in URu$_{2}$Si$_{2}$ and other systems has proven challenging, the dual character of correlated electrons, partially localized and itinerant and strong coupling between their spin, orbital and charge degrees of freedom, has been shown to be a key for stabilizing hidden orders\cite{Elgazzar2009a,Okazaki2011,Ikeda2012a,Chandra2013a,Tonegawa2014a,Bareille2014a,Kung2015a}. However, many questions have not yet been clarified; what extent a hidden order is generic, and whether or not it can occur in other systems with different types of interactions\cite{Mydosh2011a,Mydosh2020}.

Heterostructures built with two distinct correlated materials may offer a good opportunity to explore intriguing phases, thanks to the additional interaction between the layers. Proximity coupling, despite being usually weaker than the intralayer coupling, can be important in correlated heterostructures by precipitating the transition from a nearly degenerate phase to a hidden phase. One recent example is Sr$_{2}$VO$_{3}$FeAs, which is composed of iron arsenide (FeAs) and transition metal oxide layers (Fig. 1(a))\cite{Zhu2009d,Mazin2009a,Lee2009a}. The spin, orbital, and lattice degrees of freedom of 3$d$ Fe electrons in the FeAs layers are strongly entangled, so the resulting phase can be highly susceptible to additional proximity coupling to localized spins in the neighboring SrVO$_3$ layers (Fig. 1(b))\cite{Nakamura2010b,Qian2011a,Kim2015}. Here we show that an exotic hidden order phase transition occurs in Sr$_2$VO$_3$FeAs at  $T_{\rm{HO}}$ $\sim$ 150 K, without breaking any of the translational, rotational and time-reversal symmetries. This transition is described by a fully isotropic gap opening on only one of the Fermi surfaces (FSs) with a strong interlayer hopping, that is incompatible with currently recognized mechanisms of gap opening, such as Mott\cite{osm1,osm2,osm3} or density-wave transitions\cite{TaS2,NbSe2}. The resulting hidden order phase exhibits strong proximity coupling with localized V spins, in association with unusual magnetoresistance (MR) below $T_{\rm{HO}}$, and serves as a distinct parent state for superconductivity in Sr$_2$VO$_3$FeAs.

\begin{figure}
\centerline{\includegraphics[scale=1]{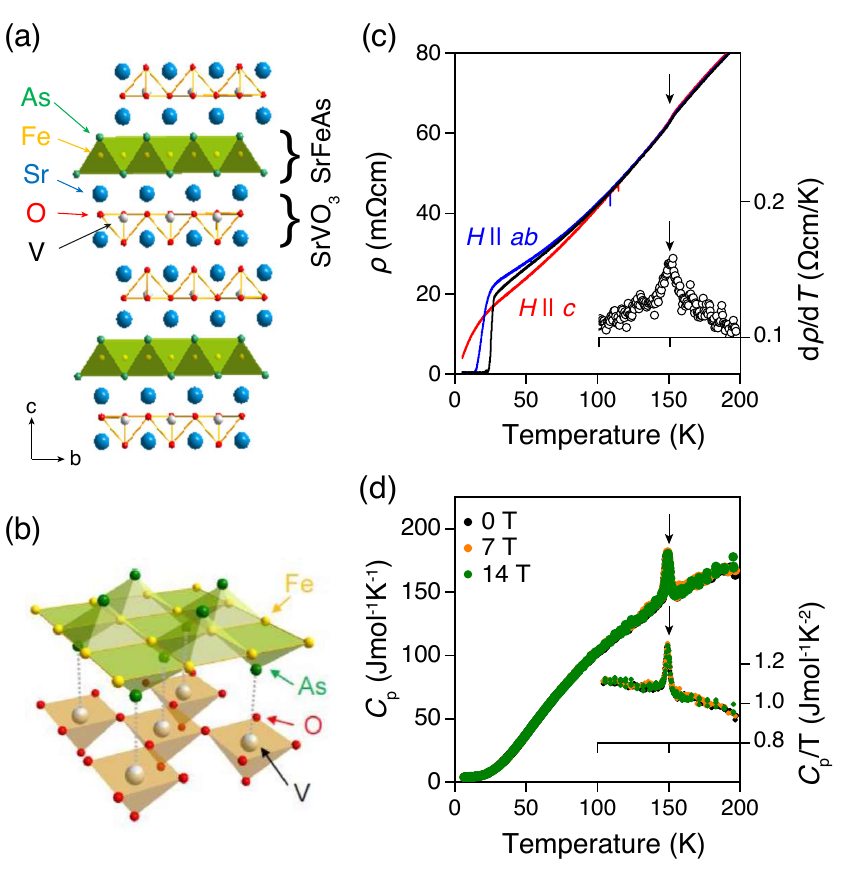}}
\caption{
Crystal structure of Sr$_2$VO$_3$FeAs and the hidden transition.
(a),(b) The crystal structure of Sr$_2$VO$_3$FeAs, a natural heterostructure with [SrFeAs]$^{+1}$ and [SrVO$_3$]$^{-1}$ layers. In the [SrVO$_3$]$^{-1}$ layers, V ions form a network of corner-sharing tetrahedrons, while the FeAs layers consist of edge-sharing FeAs$_4$ tetrahedrons. The As atoms below the center of the Fe$_4$ units of the FeAs layer are placed on top of the V atoms of the neighboring SrVO$_3$ layers, thus bridging the Fe and the V planes (indicated by the dashed lines in (b)). For clarity, the Sr atoms between these layers are removed in (b).
(c) The resistivity in the $ab$ plane ($\rho_{ab}$) shows a clear anomaly at $T_{\rm HO}$ $\sim$ 150 K, which becomes more clear in its temperature derivative $d\rho_{ab}(T)/dT$ in the inset. With a magnetic field of along $H \parallel ab$ or $H \parallel c$, no shift or broadening is observed in either $\rho_{ab}(T)$ or $d\rho_{ab}(T)/dT$ curve. The drop in resistivity below 30 K represents the superconducting transition.
(d) The specific heat $C_p$ of Sr$_2$VO$_3$FeAs as a function of temperature, at $H$ = 0, 7, 14 T. A sharp peak without magnetic field dependence is observed at $T_{\rm HO}$. The inset shows $C_p/T(T)$ near the hidden order transition at $T_{\rm HO}$.
}
\label{fig1}
\end{figure}

\section{Results}

An intriguing phase transition at $T_{\rm{HO}}$ $\sim$ 150 K is clearly indicated by the temperature-dependent resistivity $\rho(T)$ (Fig. 1(c)) and specific heat $C_p (T)$ (Fig. 1(d)) of our single crystalline samples. Neither thermal hysteresis nor magnetic field dependence in $T_{\rm{HO}}$ is observed in $\rho(T)$ up to 14 T, whereas significant MR develops below $T_{\rm{HO}}$ as we discuss below. The anomaly in $C_p (T)$ is sharper than expected in the mean-field behavior of the conventional second order transition, which may be related to a finite coupling to the $c$-axis lattice contraction below $T_{\rm{HO}}$\cite{ok}. The entropy loss across the transition $S$ $\sim$ 0.3$R$ln2, where $R$ is the gas constant, is larger than the nematic transition but smaller than the magnetic one observed in other iron-based superconductors (FeSCs)\cite{FeSecap,Ba122cap}. No evidence for broken tetragonal $C_4$ or time-reversal symmetry across $T_{\rm{HO}}$ was found in a recent nuclear magnetic resonance (NMR) study using both As and V nuclei\cite{ok}. Therefore, despite the clear thermodynamic signature of the phase transition in Fig. 1, the phase cannot be described by conventional magnetic, nematic, or density-wave orders found in FeSCs, and we thus refer to it as a hidden order\cite{Palstra1985,Maple1986a,Mydosh2011a,Elgazzar2009a,Okazaki2011,Ikeda2012a,Chandra2013a,Tonegawa2014a,Bareille2014a,Kung2015a}.

To understand the electronic response across the hidden order transition, we study the low-energy electronic structure using the angle-resolved photoemission spectroscopy (ARPES). Since the Mott insulating SrVO$_3$ layers do not contribute to the low energy states near the Fermi energy ($E_F$)\cite{Qian2011a,Kim2015}, the overall electronic structure of Sr$_2$VO$_3$FeAs shown in Figs. 2(a) and 2(b) follows the general band topology of other FeSCs\cite{FeSCAPRESreview1,FeSCnematicAPRES1,FeSCnematicAPRES2}. There are three bands centered at the $\Gamma$ point, one electron band ($\alpha$) and two hole bands ($\beta$ and $\gamma$), while two electron bands ($\delta$ and $\epsilon$) are located at the $M$ point of the Brillouin zone (BZ). For each band, the contributions from Fe orbitals (mostly $d_{xz}$, $d_{yz}$ and $d_{xy}$) differ, as summarized by the color-coded lines in Fig. 2(c), which is determined by systematic polarization-dependent ARPES studies\cite{Kim2015}. The orbital characters of these bands are similar to those of other FeSCs except for the $\gamma$ hole band. For other FeSCs, the $\gamma$ hole band near $E_F$ is derived almost entirely from the in-plane $d_{xy}$ orbital and shows a strong two-dimensional character. However, the $\gamma$ hole band of Sr$_2$VO$_3$FeAs was found to be comprised of all three $t_{2g}$ orbitals and thus has significant $k_z$ dispersion, $i.e.$, even stronger than that of the electron pockets ($\delta$ and $\epsilon$), which are mostly derived from $d_{xz,yz}$ orbitals (see Supplemental Materials Fig. S1). This unusual orbital mixing, and the resulting $k_z$ dispersion of the $\gamma$ hole band, are critical to its response across $T_{\rm HO}$. Such behavior is distinct from that of other FSs, as discussed below.

\begin{figure*}
\centerline{\includegraphics[scale=1]{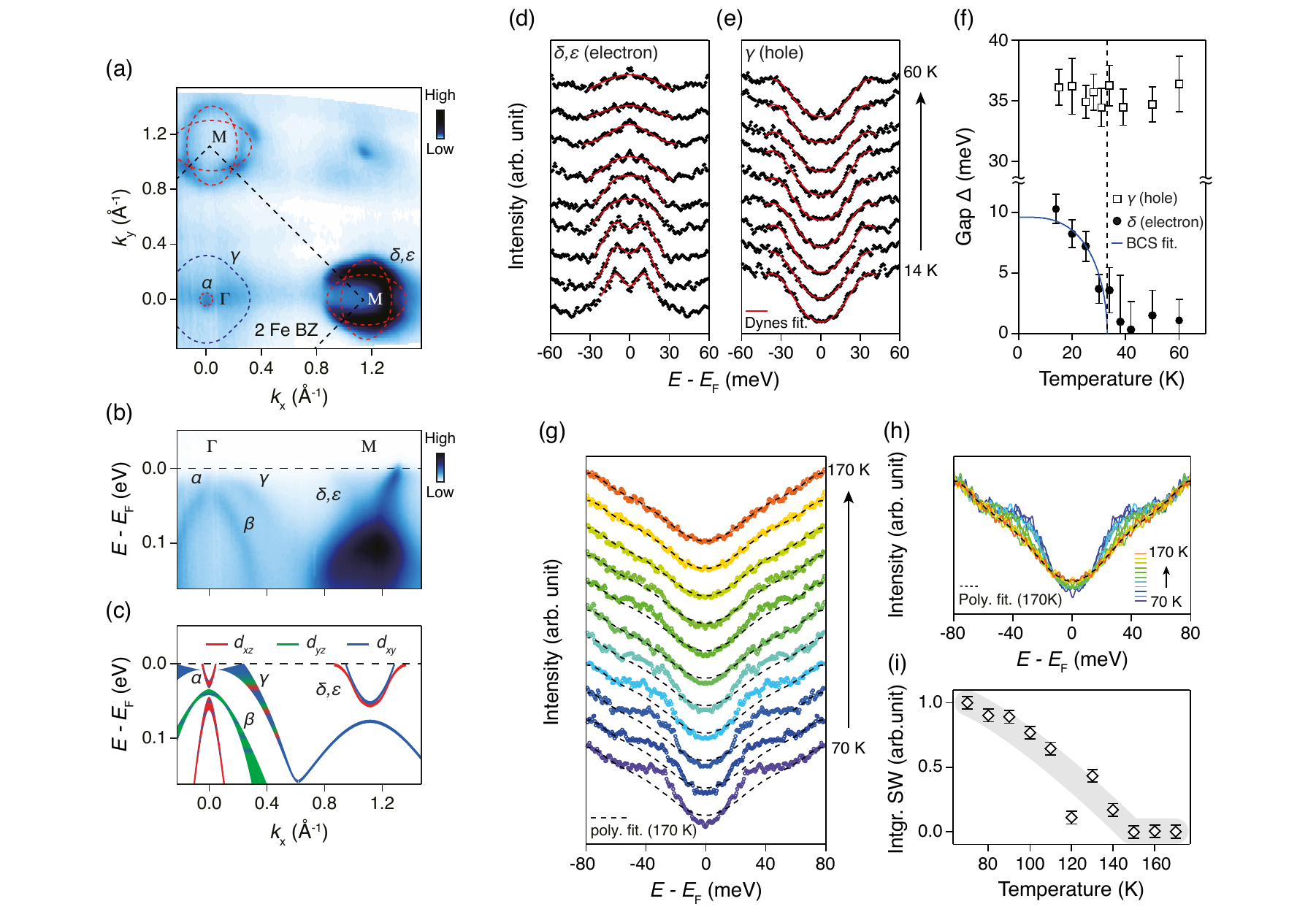}}
\caption{
Electronic structure of Sr$_2$VO$_3$FeAs and anomalous temperature-dependent gap.
(a) FS map of Sr$_2$VO$_3$FeAs observed by ARPES. Red and blue dashed lines overlaid on the figure are electron ($\alpha, \delta, \epsilon$) and hole ($\gamma$) FSs, respectively. The black dashed line represents BZ in the 2 Fe unit cell.
(b) Band structure along (0,0)-($\pi$,0) direction.
(c) Schematic of the band structure showing color coded orbital contributions ($d_{xz}$, $d_{yz}$, and $d_{xy}$).
(d),(e) Symmetrized EDCs from Fermi momenta ($k_F$) of electron ($\delta$, $\epsilon$) and hole ($\gamma$) bands, respectively, for different temperatures. The overlaid red curves are results of Dynes formula fits.
(f) Gap sizes obtained by applying the Dynes formula fit to the data in (d) and (e) as a function of temperature. The error bars include the uncertainty in the position of $E_F$.
(g) Symmetrized EDCs of the $\gamma$ hole band for temperatures between 70 and 170 K. Overlaid black dashed lines are the polynominal fit of the 170 K spectrum.
(h) The spectra in (g) are overlaid on top of each other to show the temperature evolution more clearly.
(i) Integrated spectral weight of the difference between the data and polynominal fit of the 170 K spectrum as a function of the temperature. The integrated spectral weight of 70 K is normalized to 1.
}
\label{fig2}
\end{figure*}

As well as identifying the overall band structure of Sr$_2$VO$_3$FeAs, we also investigate the detailed low-energy electronic structure near the $E_F$. We first compare the gap opening in the electron ($\delta$, $\epsilon$) and hole ($\gamma$) bands (Figs. 2(d) and 2(e)). The symmetrized energy distribution curves (EDCs) taken at $T$ = 14 K, $i.e.$, well below $T_C$ of $\sim$ 30 K, exhibit a clear gap for both $\delta$ electron and $\gamma$ hole bands. The gap sizes for $\delta$ and $\gamma$ bands, estimated with the Dynes formula, differ significantly from each other, as shown in Fig. 2(f) ($\Delta_{\delta}$ $\sim$ 10 meV and $\Delta_{\gamma}$ $\sim$ 35 meV for electron and hole bands, respectively). A gap is also observed for the small $\alpha$ electron band, and its size follows the superconducting gap function $\Delta$ = $\Delta_0$$|\cos k_x \cos k_y|$, together with those of $\delta$ bands (see Supplemental Materials Fig. S3).  The gap in the $\delta$ electron band obtained from the temperature-dependent spectra (Fig. 2(d)) follows the BCS-like temperature evolution and closes at $T_C$ (Fig. 2(f)). 2$\Delta/k_{\rm B}T_C$ is found to be $\sim$ 8, a typical value for strongly coupled superconductors.

In contrast to the gap behavior of electron bands, surprisingly, the size of the larger gap in the $\gamma$ hole band remains similar across $T_C$ (Figs. 2(e) and 2(f)). To investigate the behavior of the anomalous gap in the $\gamma$ hole band, we obtained data at higher temperatures, in the range between 70 and 170 K, as shown in Fig. 2(g). We take the polynomial fit of the spectrum at 170 K as the normal state spectrum and overlay them as dashed lines in Fig. 2(g). First, it should be noted from the 70 K data is that the spectral weight is transferred from $E_{\rm F}$ to a peak at a higher binding energy of $\sim$ 40 meV. The overlaid spectra in Fig. 2(h) clearly show that the gap is gradually filled, and the peak is suppressed as the temperature increases past $T_C$. Both the gap and peak finally disappear near $T_{\rm HO}$ $\sim$ 150 K (see Fig. 2(i) and Supplemental Materials Fig. S4). A clear ``peak-and-gap'' line shape is the key feature distinct from the so-called pseudo-gap behavior of other FeSCs\cite{FeSCPG1, FeSCPG2} and high-$T_C$ cuprates\cite{CupratePG1, CupratePG2} -- a partial gap feature only due to spectral weight transfer to a much wider energy window. We can see that the gap size has little temperature dependence, which is in contrast to the gap evolution expected in the mean-field theory, but is consistent with the sharp anomaly in the specific heat data (Fig. 1(d)). These observations unambiguously reveal that, across $T_{\rm HO}$, unusual electronic gap opening occurs in the hole Fermi pocket with a gap ratio 2$\Delta/k_{\rm B}T_{\rm HO}$ $\sim$ 6, while the electron pockets remain gapless above $T_C$.

\begin{figure*}
\centerline{\includegraphics[scale=1]{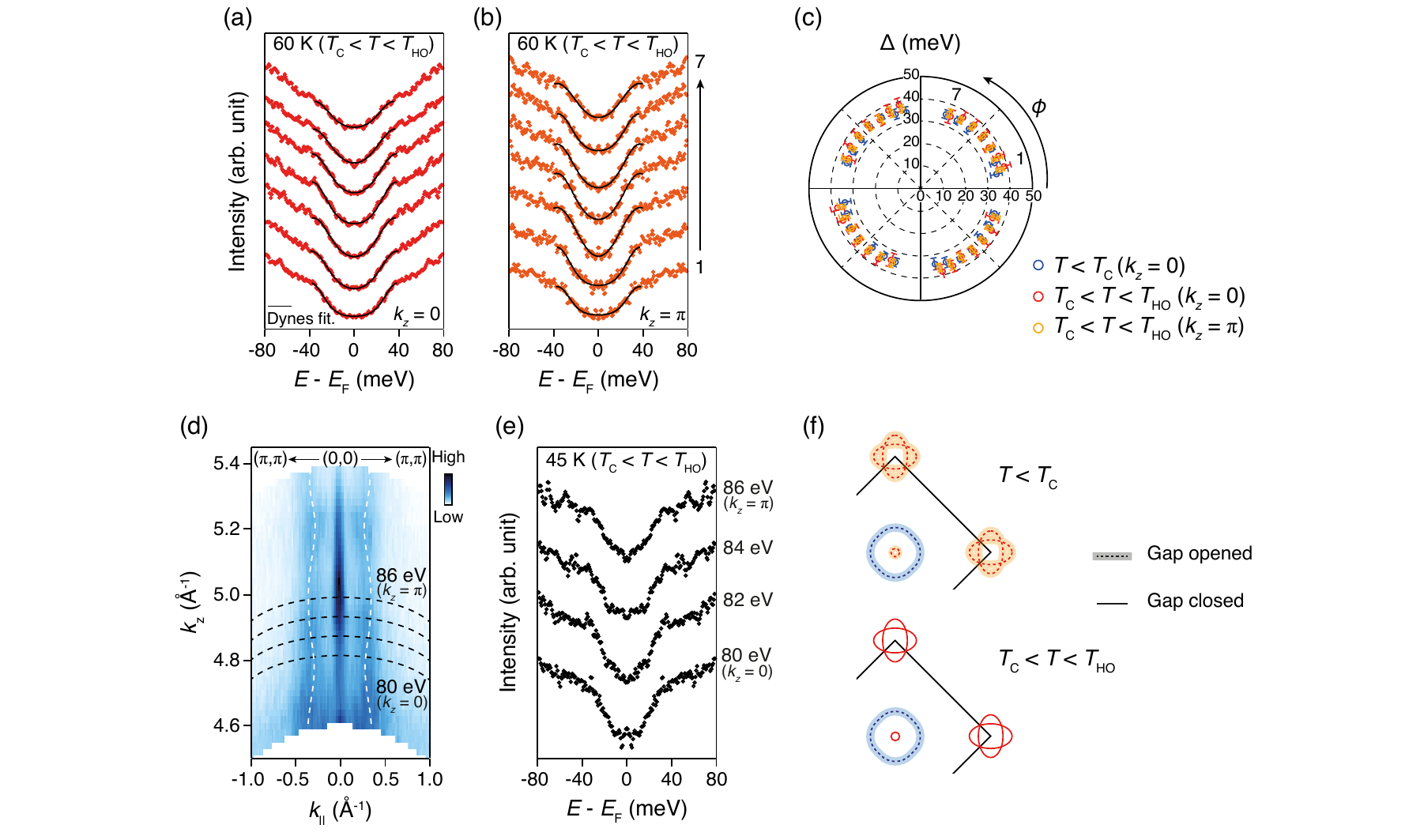}}
\caption{
Isotropic and $k_z$-independent gap in the hole band.
(a),(b) Symmetrized EDCs from different momenta on the hole FS at $k_z$ = 0 (a) and $k_z$ = $\pi$ (b) planes at $T$ = 60 K ($T_C$ $<$ $T$ $<$ $T_{\rm HO}$). Overlaid black solid curves are results of Dynes formula fits.
(c) The gap size for different $k_z$ and temperatures as a function of azimuthal angle ($\phi$) with respect to the $k_x$ axis. Only the data points in the first quadrant are actual data points; those in other quadrants are derived from symmetrization.
(d) $k_z$ band dispersion along the (0,0)-($\pi$,$\pi$) direction near the BZ center. White dashed lines are drawn as a guide for the eye for $\gamma$ hole band dispersion.
(e) Symmetrized EDCs from the hole band at different photon energies of 80, 82, 84, and 86 eV. The corresponding $k_z$ positions are denoted by black dashed lines in (d). All spectra are taken at 45 K, between $T_C$ and $T_{\rm HO}$.
(f), Schematic illustration of the gap behavior for hole and electron FSs at two different temperatures, $T$ $<$ $T_C$ (upper) and $T_C$ $<$ $T$ $<$ $T_{HO}$ (lower).
}
\label{fig3}
\end{figure*}

Having established that the unusual gap opening occurs only in the hole pocket below $T_{\rm HO}$, we now examine the in-plane momentum dependence of the gap to determine whether there is any signature of symmetry breaking in the gap structure related to the phase transition. Figure 3(a) depicts symmetrized EDCs taken along the hole FS (see Fig. 3(c)) at $T$ = 60 K ($T_C$ $<$ $T$ $<$ $T_{\rm HO}$) for $k_z=0$. Note that the peak-and-gap feature is present in all of the spectra, that is, in all directions. The gap sizes of the spectra are again estimated with the Dynes fit, as indicated by the overlaid curves in Fig. 3(a); the results are plotted in Fig. 3(c). The anomalous gap at $T_C$ $<$ $T$ $<$ $T_{\rm HO}$ is almost independent of the in-plane momentum. Assuming a possible azimuthal angle ($\phi$) dependence in the $k_x$-$k_y$ plane, expressed by $\Delta(\phi)= \Delta_0 + \Delta_n\cos n \phi$, the $n$-fold gap anisotropy $\Delta_n$ should be less than $\sim$ 5\% of isotropic gap $\Delta_0$.

It is still conceivable that the gap may not be isotropic at other $k_z$ values as there is fairly large $k_z$ dispersion for the $\gamma$ hole pocket\cite{Kim2015}; the ARPES intensity map at 10 meV in the $k_z$-$k_{\parallel}$ plane near the BZ center in Fig. 3(d) indeed clearly shows that the $\gamma$ hole pocket has $k_z$ dispersion. We therefore took the same set of $\gamma$ hole band gap data for $k_z=\pi$ and plotted these in Fig. 3(b). The peak-and-gap feature is developed below $T_{\rm HO}$ (see Supplemental Materials Fig. S5), as is the case for the symmetrized EDCs taken at $k_z=0$ (Fig. 3(a)). The procedure used to extract the gap from the $k_z=0$ data yields an isotropic gap for $k_z=\pi$, as plotted in Fig. 3(c). It should be noted that the gap sizes for $k_z=0$ and $\pi$ are almost the same, as seen in Fig. 3(c). The EDCs taken with various photon energies between $h\nu$ = 80 and 86 eV (corresponding $k_z$ values are indicated by the black dashed lines in Fig. 3(d)) are also found to have essentially the same gap size (Fig. 3(e)). The observed anomalous gap above $T_C$ is therefore independent of in-plane and out-of-plane momentum, indicating a fully isotropic gap in the $\gamma$ pocket. The fully isotropic gap confirms that the rotational and translational symmetries of the underlying crystal lattice are retained across $T_{\rm HO}$, which is consistent with the result of previous studies\cite{ok}.

The observed anomalous full gap opening cannot easily be explained by the currently recognized mechanisms of gap opening. In particular, the band-selective and fully isotropic gap opening set strong constraints on its origin. The isotropic gap in the $\gamma$ pocket, centered at the $\Gamma$ point, together with the lack of splitting of degenerate $d_{xz}$ and $d_{yz}$ states of $\delta$ and $\epsilon$ bands (Fig. 2), reflects the $C_4$ symmetry of the tetragonal lattice maintained in the FeAs layers across $T_{\rm HO}$. This is consistent with the nearly isotropic Knight shift at $T_{\rm HO}$ seen in the recent NMR study\cite{ok}, and thus rules out the $C_2$ nematic order, which is one of the most common orders in FeSCs\cite{FeSCnematicAPRES1,FeSCnematicAPRES2}. Furthermore, the almost isotropic gap over the whole $\gamma$ FS (Fig. 3(c)) implies that the translational symmetry of the underlying lattice is retained across the transition. In density-wave type transitions with a non-zero $\bm{Q}$ modulation, for example, strong gap opening occurs at sections of FS that are connected with the modulation vector $\bm{Q}$, leading to significant gap anisotropy in the momentum space\cite{TaS2,NbSe2}. As shown in our simulations based on mean-field theory for various density-wave orders, the relative gap anisotropy with respect to the isotropic gap $\Delta_n$/$\Delta_0$ is larger than $\sim $ 1 (see Supplemental Materials Fig. S10), clearly distinct from our experimental results. 

A band-selective Mott transition, which is often found in FeSCs\cite{osm1,osm2,osm3}, is also unlikely. In a  band-selective Mott transition, a particular heavy-band loses its coherent spectral weight and becomes localized due to the strong correlation effect, while the other bands remain itinerant. In this case, as found for the $d_{xy}$ bands in K$_x$Fe$_{2-y}$Se$_2$\cite{osm1,osm2}, the spectral weight is completely suppressed for the entire band in the whole BZ and is transferred to the Hubbard state located far away from $E_F$ by $\sim$ $U$, which is much larger than the band width. In Sr$_2$VO$_3$FeAs, however, the spectral transfer occurs only near $E_F$ within an energy window much smaller than the band width; the rest of the $\gamma$ band remains almost intact. Our experimental findings are therefore inconsistent with theoretical models that involve breaking of time-reversal, translational, rotational symmetries, thus confirming the hidden order below $T_{\rm HO}$.

The band selectiveness of gap opening summarized in Fig. 3(f) may further provide a clue regarding the nature of hidden order. In other FeSCs, the $\gamma$ hole pocket mainly arises from the $d_{xy}$ orbital, which is believed to have a stronger correlation effect than $d_{xz,yz}$-related bands\cite{FeSCnematicAPRES1,Kotliar2011}. In Sr$_2$VO$_3$FeAs, however, there are significant contributions from $d_{xz,yz}$ orbitals to the $\gamma$ hole pocket, which thus shows strong dispersion along the $k_z$ direction. This contrasts with the cases of $\delta,\epsilon$ electron pockets at the $M$ points, which have negligible $k_z$ warping with strong two-dimensionality. The $\alpha$ electron pocket at $\Gamma$ appears to have a weak $k_z$ dispersion compared to the $\gamma$ hole pocket. These two-dimensional FSs -- $\alpha$, $\delta$, and $\epsilon$ pockets, remain gapless below $T_{\rm HO}$ until superconductivity develops below $T_C$; that is, the full gap below $T_{\rm HO}$ selectively develops at the $\gamma$ hole pocket which has a strong $k_z$ dispersion, and thus strong interlayer hopping. This suggests that proximity coupling to the neighboring Mott layers plays an essential role in triggering the hidden order.

\begin{figure*}
\centerline{\includegraphics[scale=1]{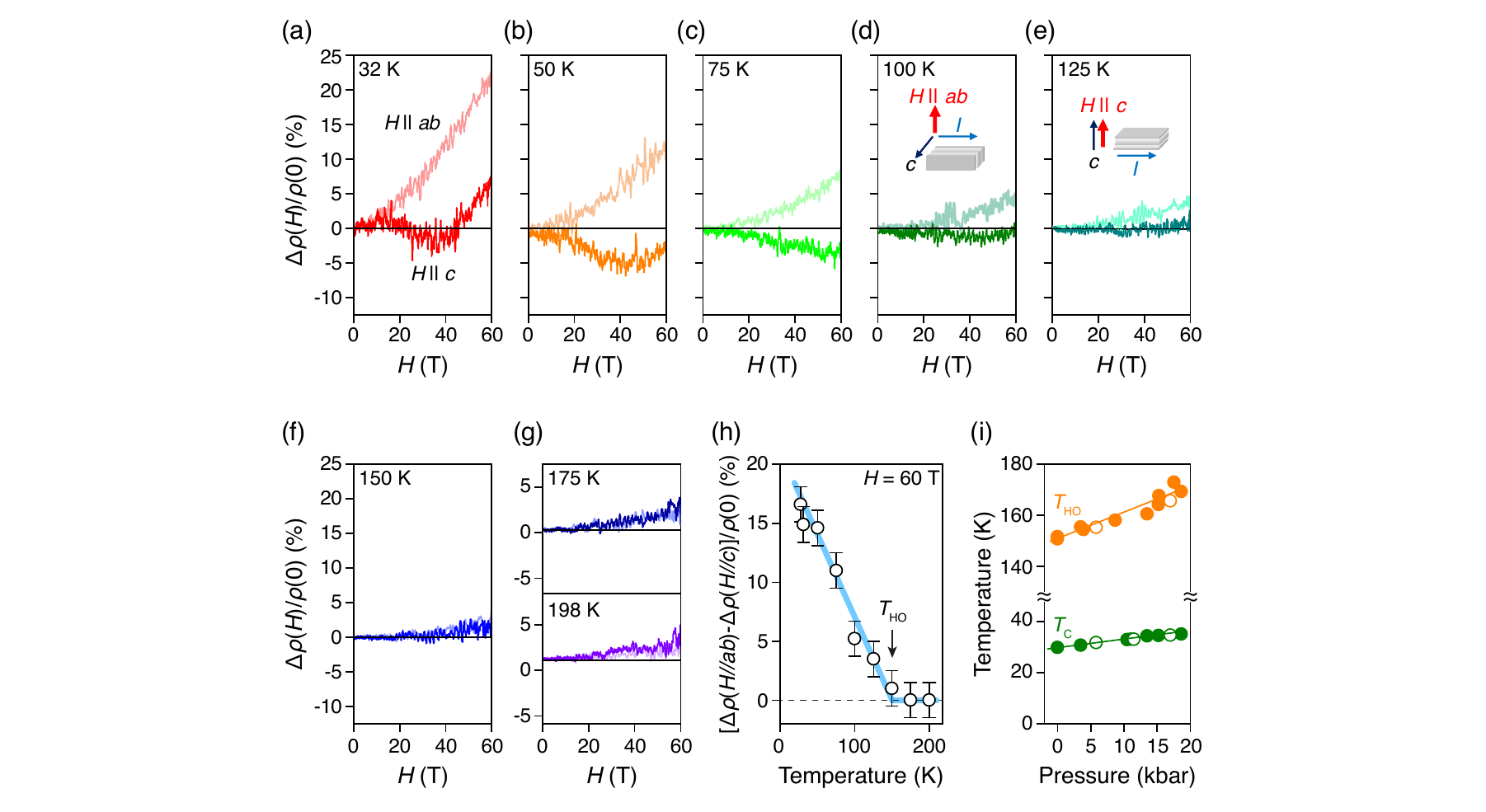}}
\caption{
Negative MR and pressure tuning.
(a)--(g) Magnetic field-dependent MR $\Delta\rho(H)/\rho(0)$ of Sr$_2$VO$_3$FeAs up to 60 T, at various temperatures for $H \parallel ab$ and $H\parallel c$. The inset shows a schematic of current and field orientations with respect to the $c$-axis. The negative MR for $H\parallel c$, which developed below $T_{\rm HO}$, is distinct from the strong positive MR for $H\parallel ab$.
(h) The anisotropy of the MR, given by difference between $\Delta\rho (H\parallel ab)$ and $\Delta \rho (H\parallel c)$, normalized by $\rho(0)$ as a function of temperature. The onset of anisotropic MR well matches $T_{\rm HO}$. 
(i) Pressure dependence of the hidden order ($T_{\rm HO}$) and the superconducting ($T_C$) temperatures.
}
\label{fig4}
\end{figure*}

The importance of the proximity coupling is reflected in the unusual transport properties seen under magnetic fields. Figure 4 shows the in-plane MR, $\rho_{ab}(H)$, under magnetic fields up to 60 T for $H \parallel ab$ and $H \parallel c$ in the transverse configuration. At high temperatures above $T_{\rm HO}$, positive MR is seen for both $H \parallel c$ and $H \parallel ab$, with that for $H \parallel c$ being slightly larger. This can be understood in terms of the conventional orbital effect typically found in quasi-two dimensional systems (see Supplemental Materials Fig. S6). However, an additional contribution of the negative MR starts to emerge for $H \parallel c$ below $T_{\rm HO}$, and dominates the field dependence until it is saturated at higher magnetic fields. Taking the MR for $H \parallel ab$ as the reference, we estimate the negative contribution of the MR for $H \parallel c$ by $\Delta \rho_{ab}$ = $\rho_{ab}(H\parallel ab)-\rho_{ab}(H\parallel c)$ at $H$ = 60 T (Fig. 4(h)); a clear onset at $T_{\rm HO}$ can be seen. Such negative MR has not been observed in FeSCs, except EuFe$_2$As$_2$\cite{Eu122}, in which the spin scattering due to localized Eu spins above $T_N$ is important. The observed complex field-dependent MR in Sr$_2$VO$_3$FeAs below $T_{\rm HO}$ can only be understood by taking into account the significant scattering of itinerant electrons with fluctuating localized spins, which is suppressed under high magnetic fields. This shows that the hidden order transition in Sr$_2$VO$_3$FeAs is intimately tied to the hybridization between the itinerant Fe electron fluid and localized V spins caused by proximity coupling.

In Sr$_2$VO$_3$FeAs, the localized V spins are known to remain fluctuating at least down to 5 K, even with a large Curie-Weiss temperature ($\Theta$ $\sim$ 100 K)\cite{ok,cao,tatematsu}. The absence of magnetic order for V spins in the square lattice by itself evidences additional Kondo-like coupling $J_K$ with Fe electrons across the interface, which is frustrated with intralayer superexchange interaction $J_{ex}$ of localized V spins\cite{ok}. This is consistent with the enhancement of the interlayer $J_K$ interaction across $T_{\rm HO}$ observed in the MR behavior in Fig 4. A significant spin interaction between the FeAs and the SrVO$_3$ layers is also captured in recent measurements of the spin relaxation time, which showed a spin-gap-like behavior with unusual suppression of the stripe-type antiferromagnetic (AFM) fluctuation\cite{ok}. Furthermore, we found that $T_{\rm HO}$ systematically increases with external pressure (see Fig. 4(i) and Supplemental Materials Fig. S7). External pressure mostly reduces the interlayer distance and thus enhances proximity coupling in layered compounds like Sr$_2$VO$_3$FeAs. This pressure dependence is consistent with the notion that the hybridization between itinerant electrons in the FeAs layers and the localized spins in the SrVO$_3$ layers is responsible for stabilizing the hidden order.

\section{Discussion}
Based on these results, we argue that Sr$_2$VO$_3$FeAs is another rare material that hosts a hidden order phase following URu$_2$Si$_2$. The two systems share common features, such as a coupling between localized spins and itinerant electrons, a clear specific heat anomaly and electronic gap opening at $T_{\rm HO}$, and coexistence with superconductivity and antiferromagnetism\cite{Mydosh2011a}.
However, we stress that the nature of coupling in Sr$_2$VO$_3$FeAs differs from that in URu$_2$Si$_2$.
Namely, itinerant Fe 3$d$ electrons in Sr$_2$VO$_3$FeAs are spatially separated from the localized V spins and coupled through As orbitals (Fig. 1(b)), whereas the uranium 5$f$ electrons in URu$_2$Si$_2$ are on the verge of being localized and itinerant with the strong on-site Kondo coupling, which is considered as important ingredient for stabilizing the hidden order\cite{Mydosh2011a,Mydosh2020}. The Kondo-like proximity coupling $J_K$ in Sr$_2$VO$_3$FeAs is essentially non-local, which can introduce strong momentum-space anisotropy in hybridization\cite{Senthil, Vojta}. Furthermore, itinerant Fe electrons have internal orbital degree of freedom, particularly in the $\gamma$ band with highly mixed 3$d_{xz,yz}$ and 3$d_{xy}$ orbitals. These non-local and multiorbital features may play a role in the abrupt change in $J_K$ across $T_{\rm HO}$, which could lead to the band selective gap opening (Fig. 2), the onset of negative MR (Fig. 4), and the reduction of the $c$-axis lattice parameter\cite{ok}. This would also significantly affect the superconducting order in the FeAs layers, by suppressing typical ($\pi$,0) spin fluctuations between the $\gamma$ hole and the $\delta,\epsilon$ electron FSs and invoking other pairing channels, $e.g.$, with the incipient $\beta$ hole bands\cite{bang,Hirschfeld} or via other $C_4$ symmetric AFM fluctuations\cite{vestigial}. The possibility of more exotic scenarios associated with deconfined phases, such as orthogonal metals\cite{Nandkishore2012,Sachdev,egmoon} for hidden order and the coexistence of a superconducting order, remains to be explored. Our findings regarding the hidden order of Sr$_2$VO$_3$FeAs highlight that interfacing 3$d$ transition pnictides and oxides could host diverse exotic phases with entangled spin, orbital, and charge degrees of freedom in tunable manner through proximity coupling.

\begin{acknowledgments}
The authors thank I. Mazin, S.-H. Baek, H. W. Yeom, and Y. Bang for fruitful discussion.
The work at Korea Advanced Institute of Science and Technology (KAIST) is supported by the National Research Foundation of Korea (NRF) with grant No.2018R1D1A1B07050869, 2018K1A3A7A09056310, 2015M3D1A1070672, 2017R1C1B2009176, and 2019M3E4A1080411. The work at Pohang University of Science and Technology (POSTECH) was supported by the Institute for Basic Science (IBS) through the Center for Artificial Low Dimensional Electronic System (No. IBS-R014-D1), and by the NRF through SRC (Grant No. 2018R1A5A6075964), the Max Planck-POSTECH Center for Complex Phase Materials (Grant No. 2016K1A4A4A01922028). The work at Seoul National University (SNU) was supported by the research program of the IBS (Grant No. IBS-R009-G2). The Advanced Light Source is supported by the Office of Basic Energy Science of the U.S. DOE under contract No. DeAC02-05CH11231.
\end{acknowledgments}

\appendix
\section{Methods}
\subsection{Single crystal growth}
Sr$_{2}$VO$_{3}$FeAs single crystals were grown using self-flux techniques as follows. The mixture of SrO, VO$_{3}$, Fe, SrAs, and FeAs powders with a stoichiometry of Sr$_{2}$VO$_{3}$FeAs:FeAs = 1:2 were pressed into a pellet and sealed in an evacuated quartz tube under Ar atmosphere. The samples were heated to 1180$^{o}$C, held at this temperature for 80 hours, cooled slowly first to 950$^{o}$C at a rate of 2$^{o}$C/h and then furnace-cooled. The plate-shaped single crystals were mechanically extracted from the flux. High crystallinity and stoichiometry are confirmed by the X-ray diffraction and energy-dispersive spectroscopy. The typical size of the single crystals is 200$\times$200$\times$10 $\mu$m$^{3}$.\\

\subsection{Transport properties and specific heat}
Magnetotransport properties were measured using conventional four-probe configuration on a single crystal in a 14 T Physical Property Measurement System (Quantum Design), a 33 T Bitter magnet at the National High Magnetic Field Lab., Tallahassee, and a 60 T pulse magnet at Dresden High Magnetic Field Lab., Dresden. Specific heat measurements were done on several pieces of Sr$_{2}$VO$_{3}$FeAs\ single crystals ($\sim$ 1 mg) using the relaxation method in a 14 T Physical Property Measurement System (Quantum Design).\\

\subsection{Angle resolved photoemission spectroscopy}
ARPES measurements were performed at beam lines 10.0.1 (HERS) and 4.0.3 (MERLIN) of the Advanced Light Source, Lawrence Berkeley National Laboratory. Samples were cleaved at 10 K in ultra-high vacuum better than 3 $\times$ 10$^{11}$ Torr. Spectra were acquired with Scienta R4000 analyzer at 10.0.1 and Scienta R8000 analyzer at 4.0.3. Several photon energies, particularly, between 40 and 70 eV at beam line 10.0.1, and between 68 and 100 eV at beam line 4.0.3, were used for the ARPES measurements including photon energy dependence. The overall energy resolution was 18 meV or better. 


\end{document}


\preprint{}

\title{Supplemental Material:
Proximity-induced hidden order transition in a correlated heterostructure Sr$_2$VO$_3$FeAs}

\author{Sunghun Kim}
\thanks{These authors contributed equally to this work}
\affiliation{Department of Physics, Korea Advanced Institute of Science and Technology, Daejeon 34141, Republic of Korea}

\author{Jong Mok Ok}
\thanks{These authors contributed equally to this work}
\affiliation{Center for Artificial Low Dimensional Electronic Systems, Institute for Basic Science, Pohang 37673, Republic of Korea}
\affiliation{Department of Physics, Pohang University of Science and Technology, Pohang 37673, Republic of Korea}

\author{Hanbit Oh}
\affiliation{Department of Physics, Korea Advanced Institute of Science and Technology, Daejeon 34141, Republic of Korea}

\author{Chang-il Kwon}
\affiliation{Center for Artificial Low Dimensional Electronic Systems, Institute for Basic Science, Pohang 37673, Republic of Korea}
\affiliation{Department of Physics, Pohang University of Science and Technology, Pohang 37673, Republic of Korea}

\author{Y. Zhang}
\affiliation{National Laboratory of Solid State Microstructures, School of Physics, Collaborative Innovation Center of Advanced Microstructures, Nanjing University, Nanjing 210093, China}
\affiliation{Advanced Light Source, Lawrence Berkeley National Laboratory, Berkeley, CA 94720, USA}

\author{J. D. Denlinger}
\affiliation{Advanced Light Source, Lawrence Berkeley National Laboratory, Berkeley, CA 94720, USA}

\author{S.-K. Mo}
\affiliation{Advanced Light Source, Lawrence Berkeley National Laboratory, Berkeley, CA 94720, USA}

\author{F. Wolff-Fabris}
\affiliation{Dresden High Magnetic Field Laboratory, Helmholtz-Zentrum Dresden-Rossendorf, Dresden, D-01314, Germany}

\author{E. Kampert}
\affiliation{Dresden High Magnetic Field Laboratory, Helmholtz-Zentrum Dresden-Rossendorf, Dresden, D-01314, Germany}

\author{Eun-Gook Moon}
\email[]{egmoon@kaist.ac.kr}
\affiliation{Department of Physics, Korea Advanced Institute of Science and Technology, Daejeon 34141, Republic of Korea}

\author{C. Kim}
\email[]{changyoung@snu.ac.kr}
\affiliation{Center for Correlated Electron Systems, Institute for Basic Science, Seoul 08826, Republic of Korea}
\affiliation{Department of Physics and Astronomy, Seoul National University, Seoul 08826, Republic of Korea}

\author{Jun Sung Kim}
\email[]{js.kim@postech.ac.kr}
\affiliation{Center for Artificial Low Dimensional Electronic Systems, Institute for Basic Science, Pohang 37673, Republic of Korea}
\affiliation{Department of Physics, Pohang University of Science and Technology, Pohang 37673, Republic of Korea}

\author{Y. K. Kim}
\email[]{yeongkwan@kaist.ac.kr}
\affiliation{Department of Physics, Korea Advanced Institute of Science and Technology, Daejeon 34141, Republic of Korea}
\affiliation{Graduate School of Nanoscience and Technology, Korea Advanced Institute of Science and Technology, Daejeon 34141, Republic of Korea}

\date{\today}
\maketitle
\renewcommand{\theequation}{\arabic{equation}}


\section{Angle-resolved photoemission spectroscopy}
\subsection{$k_z$ dependence of $\delta$ electron band}
Figure S1(a) shows ARPES intensity map of electron band near $M$ point at $E_F$ in $k_z$-$k_x$ plane. It is clearly demonstrated that $k_F$ are almost constant along $k_z$ direction, shown with overlaid white dashed guidelines. Stack of selected momentum distribution curves (MDCs) extracted from Fig. S1(a) are plotted in Fig. S1(b). It indeed shows negligible modulation of peak position along $k_z$ direction.
Figure S1(c) shows $k_F$ of $\gamma$ hole band (blue) and $\delta$ electron band (red) at several different $k_z$. As described above, $k_F$ of $\delta$ electron band are almost constant at different $k_z$ while the $\gamma$ hole band shows finite modulation along $k_z$ direction (Fig. 3(d) of main text and Fig. S1(c)). It evidences that $\gamma$ hole band has relatively strong $k_z$ dispersion compare to other electron bands.

\subsection{Charge neutrality of the Sr$_2$VO$_3$FeAs single crystal}
Charge neutrality of the system was checked through accounting the FS size, using Luttinger's sum rule. Figure S2(a) displays schematic of FS of Sr$_2$VO$_3$FeAs with 2 Fe unit cell BZ extracted from the experimental data. Electron ($\alpha$, $\delta$, $\epsilon$ pockets) and hole ($\gamma$ pocket) pocket sizes $S$ were estimated as $S_{\alpha}$ $\sim$ 0.01 \AA$^{-2}$, $S_{\delta,\epsilon}$ $\sim$ 0.17 \AA$^{-2}$, and $S_{\gamma}$ $\sim$ 0.32 \AA$^{-2}$, respectively. The average of $S_{\gamma}$($k_z$ = 0) $\sim$ 0.29 \AA$^{-2}$ and $S_{\gamma}$($k_z$ = $\pi$) $\sim$ 0.36 \AA$^{-2}$ was taken for the $\gamma$ hole pocket size because of the $k_z$ dispersion of $\gamma$ hole band. The extracted value of the difference between electron and hole FS size, $S_{e-h}$ = (electron FS size) $-$ (hole FS size), is $\sim$ 1\% of entire BZ size, which indicates the system to be almost charge neutral. It evidences that Sr$_2$VO$_3$FeAs single crystals used in present study is free from the oxygen deficiency issue.

\subsection{Superconducting gap size}
To estimate superconducting gap opening on $\alpha$ electron band, we obtained symmetrized EDCs of $\alpha$ electron band at two different temperatures, 15 K (below $T_C$) and 60 K (above $T_C$), as plotted in Fig. S3(a). It clearly distinguishes two spectra by gap opening. At low temperature (15 K), it exhibits clear gap feature, while spectrum of high temperature (60 K) shows no evidence of gap opening. It mimics the temperature dependent behavior of $\delta$ electron band that the superconducting gap closes above $T_C$ as shown in Fig. 2(d) of the main text.

Figure S3(b) summarizes the gap size of each band, $\alpha$ (yellow), $\delta$ (red) electron bands and $\gamma$ hole band (blue), obtained below $T_C$ as a function of $|\cos{k_x}\cos{k_y}|$. We note that the gap size of electron bands follow the superconducting gap function $\Delta$ = $\Delta_0|\cos{k_x}\cos{k_y}|$ (black dashed line in Fig. S3(c)), which was reported as FS size dependent superconducting gap\cite{Nakayama,Umezawa}.
In contrast to the electron bands, the gap size of the $\gamma$ (blue) hole band distinctly deviates from this behavior which implies a non-superconducting origin of the unusual gap opening on the hole band.

\subsection{$\gamma$ hole band dispersion}
Figures S4(a) and S4(b) show the hole band dispersion around the BZ center obtained at different $k_z$ and temperatures. At low temperature (left panels), both ARPES results taken at $k_z$ = 0 and $k_z$ = $\pi$ show gap on the $\gamma$ hole band. On the other hand, at high temperature (right panels), the hole band crosses the Fermi level with \textit {finite} spectral weight. Estimated band dispersion is marked by the red dashed curve for each $k_z$ (right panels of Figs. S4(a) and S4(b)) which also indicates the hole band maximum is located above the Fermi level. The expanded view of the $k_z$ = $\pi$ data taken at 45 (left) and 170 K (right) are shown in Fig. S4(c). Peak positions (black filled circles) are obtained by MDC fitting and overlaid on the data as shown in the figures. It clearly shows that the hole band crosses the Fermi level at 170 K (above $T_{\rm HO}$), while the band does not disperse through the Fermi energy with a gap for 45 K data (below $T_{\rm HO}$ but higher than $T_C$).

\subsection{Temperature dependence of hole band gap at $k_z$ = $\pi$ plane}
In order to confirm the temperature dependence of the $\gamma$ hole band gap at different $k_z$ plane, we obtained EDCs taken with $h\nu$ = 86 eV (which is corresponding to $k_z$ = $\pi$), at various temperatures. Fig. S5(a) displays symmetrized EDCs taken at various different temperatures. It shows that the gap opened at 45 K (above $T_C$) and gradually filled up with increasing the system temperature. It finally closes above 150 K ($T_{\rm HO}$). The polynomial fit curve of the 170 K spectrum overlaid on each EDC taken at different temperatures visualize peak-and-gap feature from the hole band gap. Indeed, the integrated spectral weight exhibited in Fig. S5(b) well shows the hole band gap finally closes at 150 K. This temperature dependent behavior is consistent with that observed at $k_z$ = 0 plane (Figs. 2(g)-(i) of the main text).

\section{Transport properties}
\subsection{Angle dependent magnetoresistance}
Figure S6 presents MR as a function of magnetic field angle ($\theta$), taken at different temperatures under magnetic field of 14 T. The tilting angle $\theta$ is defined with respect to the $c$ axis and the current direction is kept to be normal to the magnetic field. Consistent with the higher field data in the main text (Fig. 4), the negative contribution to MR for $H \parallel c$ develops below $T_{\rm HO}$, which produces the opposite angle dependence of MR across $T_{\rm HO}$. This contrasts to the typical behavior of the layered normal metals in which the MR for $H \parallel c$ is larger than for $H \parallel ab$. The difference between MR data taken at $\theta$ = 0 and $\pi/2$ clearly exhibit the onset at $T_{\rm HO}$.

\subsection{Pressure dependence}
The importance of the interlayer coupling to the hidden order is confirmed by the pressure effect on $T_{\rm HO}$. In the layered compounds, like Sr$_2$VO$_3$FeAs, the external pressure mostly reduces the interlayer distance, while keeping the in-plane lattice constant nearly intact. Figure S7 shows temperature dependence of the in-plane resistivity $\rho(T)$ at different pressures. The kink in $\rho_{ab}(T)$ curves, which is more clearly visible in the temperature-derivative curves $d\rho_{ab}(T)/dT$ as shown in the inset, is used to estimate $T_{\rm HO}$ at different pressures. Upon increasing pressure, $T_{\rm HO}$ is systematically shifts toward higher temperatures (Fig. S7) with a rate of $dT_{\rm HO}/dP$ $\sim$ 1 K/kbar. We found consistent pressure dependence of $T_{\rm HO}$, taken from two different crystals, independent of pressure sweep directions. These findings support that the key physics of the hidden order is the formation of an itinerant Fe electron fluid, strongly hybridized with localized V spins through interlayer interaction.

\section{Symmetry analysis and gap opening conditions}
\subsection{Model}
We consider a tight-binding model with two orbitals, $3d_{xz},3d_{yz}$, of $\mathrm{Fe}$, motivated by Refs. \cite{Raghu,FuchunZhang},
	\begin{eqnarray}
 H_{0}	&= & \sum_{\bm{k},s}\psi_{\bm{k},s}^{\dagger}\left( (\epsilon_{0}(\bm{k})-\mu)I_{2}+\epsilon_{1}(\bm{k})\tau_{1}+\epsilon_{3}(\bm{k})\tau_{3}\right) \psi_{\bm{k},s}, \label{e1}
	\end{eqnarray}	
with the Pauli matrices $\tau_{a}$, hopping parameters $t_{b}$ (for $a=0,1,3$, $b=1,\cdots 4$),  and
	\vspace{-5pt}
	\begin{eqnarray}
	&&\epsilon_{1}(\bm{k})=-4t_{4} \sin k_{x} \sin k_{y},\ \ \epsilon_{0/3}(\bm{k})=\frac{\epsilon_{x}(\bm{k})\pm \epsilon_{y}(\bm{k})}{2},\nonumber \\
	&&\epsilon_{x}(\bm{k})=-2t_{1}\cos k_{x}-2t_{2} \cos k_{y}-4t_{3} \cos k_{x} \cos k_{y} ,\label{e2}\\
	&&\epsilon_{y}(\bm{k})=-2t_{2} \cos k_{x}-2t_{1} \cos k_{y}-4t_{3} \cos k_{x} \cos k_{y}. ,\nonumber
	\end{eqnarray}
	 We consider a two dimensional system with a momentum $\bm{k}=(k_x, k_y)$ for simplicity. A two-component spinor, $\psi_{\bm{k},s}^{T}=(c_{\bm{k},xz,s},c_{\bm{k},yz,s})$, is introduced with an annihilation operator ($c_{\bm{k},\alpha,s}$) of an electron with an orbital ($\alpha=d_{xz},d_{yz}$) and a spin ($s=\uparrow,\downarrow$). Our  model may be considered as a minimal model to describe $\mathrm{FeAs}$, and its generalization to include $d_{xy}$ orbital and $z$ directional momentum is straightforward.
In Fig. S8, we illustrate a typical FS shape of the model.

\subsection{Order parameters with translational symmetry}
We first present symmetry classification. The point group of our system is $G=D_{4}\times \mathcal{T} \times \mathrm{SU(2)}\times \mathrm{U(1)}$ where $\mathcal{T}$ and $\mathrm{SU(2)}$ are time-reversal and spin-rotational symmetry, respectively. An order parameter of the point group, $\phi_{R}$, is coupled to a fermion bilinear operator,
 \begin{eqnarray*}
 \delta H_{R} &=&\phi_{R} \sum_{\bm{k}}\Psi_{\bm{k}}^{\dagger}\left( h^{ij}(\bm{k})\tau_{i} \otimes \sigma_{j}\right)\Psi_{\bm{k}},
 \end{eqnarray*}
with a four-component spinor $\Psi_{k}^{T}=(c_{\bm{k},xz,\uparrow},c_{\bm{k},yz,\uparrow},c_{\bm{k},xz,\downarrow},c_{\bm{k},yz,\downarrow})$. The subscript $R$ specifies a representation of $G$ and $\tau_{i},\sigma_{i}$ are Pauli-matrices acting on an orbital and spin space. Note that a spin-orbit coupling is ignored in our analysis since it is negligible. We identify symmetry actions on the matrix components with a fermion operator $\Psi_{\bm{k}}$ in Table S1.

With the symmetry actions in hand, we now classify the order parameters.
The spin-rotational symmetry gives two types, spin-independent ($\phi_{o}$) and dependent ($\phi_{s}$) order parameters.
	\begin{eqnarray}
 \delta H_{o} &=&\phi_{o}^{\Gamma,\mathcal{T}} \sum_{\bm{k}}\Psi_{\bm{k}}^{\dagger}\ M^{\Gamma,\mathcal{T}}_{o}(\bm{k}) \otimes \sigma^{0}\Psi_{\bm{k}}\\
  \delta H_{s} &=&\vec{\phi}_{s}^{\Gamma,\mathcal{T}}\cdot \sum_{\bm{k}}\Psi_{\bm{k}}^{\dagger}\  M^{\Gamma,\mathcal{T}}_{s}(\bm{k}) \otimes \vec{\sigma}\ \Psi_{\bm{k}}
	\end{eqnarray}
The superscripts $\Gamma, \mathcal{T}$ are for irreducible representations of a point group $D_{4}$ and a time-reversal symmetry, and the subscripts $o,s$ are for the spin-rotational symmetry. The explicit form of a $2\times 2$ matrix $M^{\Gamma,\mathcal{T}}(\bm{k})$ is listed in Table S2.

\subsection{Gap opening}
We demonstrate that a full gap opening on a FS is impossible unless superconductivity appears. To analyze a gap structure, we search for zeros of the determinant of a Hamiltonian density, $ \mathcal{H}(\bm{k})$, assuming that quasiparticles are well-defined. An original FS $(\bm{k}_{\mathrm{FS}})$ without breaking any symmetries may be obtained by the condition, $\mathrm{Det}\big( \mathcal{H}_{0}(\bm{k}_{\mathrm{FS}})\big)=0$, with
\begin{eqnarray}
\mathrm{Det}\big( \mathcal{H}_{0}(\bm{k})\big)&=&\left(\epsilon_{0}(\bm{k})^{2}-\vec{\epsilon}(\bm{k})^{2}\right)^{2}=\left(\langle \underline{\epsilon}(\bm{k}), \underline{\epsilon} (\bm{k})\rangle\right)^2\geq 0.
\end{eqnarray}
The tight-binding parameters ($\epsilon_{a}(\bm{k})$) in Eq.(\ref{e2}) are used, and  a four-vector notations $\underline{\epsilon}=(\epsilon_{0},\vec{\epsilon})$ and an inner product $\langle \underline{a},\underline{b}\rangle =a_{0}b_{0}-\vec{a}\cdot \vec{b}$ are introduced.

The shape of FS is modified by the onset of an order parameter. The total Hamiltonian density becomes $\mathcal{H}=\mathcal{H}_{0}+g\delta \mathcal{H}_{o,s}$  with a small parameter $g$ near the onset.
Its determinant  at $\bm{k}$ may be written as
	\begin{eqnarray}
	\mathrm{Det}\big( \mathcal{H}_{0}(\bm{k})+ g \delta  \mathcal{H}(\bm{k})\big)&=& \mathrm{Det}\big( \mathcal{H}_{0}(\bm{k})\big)+\sum_{n=1}^{4}g^{n}f_{n}(\bm{k}).
	\end{eqnarray}
 Note that the highest order of $g$ at each $\bm{k}$ is four due to the spinor structure.
  The $n$-th order functions, $f_{n}(\bm{k})$, depend on the parameters ($\epsilon_{a}(\bm{k})$, $m^{\Gamma,\mathcal{T}}(\bm{k})$), where $M^{\Gamma,\mathcal{T}}(\bm{k})=\sum_{a}m^{\Gamma,\mathcal{T}}_{a}(\bm{k})\tau_{a}$.

Near the onset ($g \ll1$), the lowest order non-zero $f_{n}(\bm{k})$ determines a criteria of the gap opening.

 For a spin-independent order parameter, we find that a non-zero leading order term is
  \begin{eqnarray}
g f_{1}(\bm{k})&=
& g \sum_{\Gamma,\mathcal{T}}\ 4\phi_{o}^{\Gamma,\mathcal{T}} \langle \underline{\epsilon}(\bm{k}), \underline{\epsilon} (\bm{k})\rangle\langle \underline{\epsilon}(\bm{k}), \underline{m}_{o}^{\Gamma,\mathcal{T}}(\bm{k}) \rangle.
  \end{eqnarray}
  The original FS condition, $\mathrm{Det}\big( \mathcal{H}_{0}(\bm{k}_{\mathrm{FS}})\big)=0$, gives $f_{1}(\bm{k}_{F})=0$, and we find that the signs of $f_{1}(\bm{k})$ inside and outside the FS are opposite. Hence, the original FS is shifted by the contribution $g f_1(\bm{k})$,
  \begin{eqnarray*}
  \bm{k}_{\mathrm{FS}}&\rightarrow&  \bm{k}_{\mathrm{FS}}+\delta \bm{k}_{\perp},\ \ \  |\delta \bm{k}_{\perp}|=-2g \left(\partial_{\perp}f_{1}/\partial^{2}_{\perp} \mathrm{Det}(\mathcal{H}_{0}) \right)\rfloor_{\bm{k}=\bm{k}_{\mathrm{FS}}},
  \end{eqnarray*}
   where $\bm{k}_{\perp}$ is a normal vector to the FS and $\partial_{\perp}$ is a derivative with respect to the $\bm{k}_{\perp}$.

   For a spin-dependent case, the first order term in $g$ is zero for all momentum space, and the second order term is
   \begin{eqnarray}
g^2 f_{2}(\bm{k})&=&g^2 \sum_{\Gamma,\mathcal{T}}\ 2|\vec{\phi}_{s}^{\ \Gamma,\mathcal{T}}|^{2}\left[ \langle \underline{\epsilon}(\bm{k}), \underline{\epsilon} (\bm{k})\rangle  \langle \underline{m}_{s}^{\Gamma,\mathcal{T}}(\bm{k}), \underline{m}_{s}^{\Gamma,\mathcal{T}}(\bm{k})\rangle
-2 \left(\langle \underline{\epsilon}(\bm{k}), \underline{m}_{s}^{\Gamma,\mathcal{T}}(\bm{k}) \rangle\right)^{2} \right].
   \end{eqnarray}
The second term on the right hand side gives $f_{2}(\bm{k}_{\mathrm{FS}})\leq 0$, and thus the original FS is shifted by the contribution $g^2 f_2(\bm{k})$,
   \begin{eqnarray*}
  \bm{k}_{\mathrm{FS}}&\rightarrow&  \bm{k}_{\mathrm{FS}}+\delta \bm{k}_{\perp},\ \ \  |\delta \bm{k}_{\perp}|=\pm |g| \left(-2f_{2}/\partial^{2}_{\perp} \mathrm{Det}(\mathcal{H}_{0}) \right)^{1/2}\rfloor_{\bm{k}=\bm{k}_{\mathrm{FS}}}.
  \end{eqnarray*}
   Note that $\partial^{2}_{\perp} \mathrm{Det}(\mathcal{H}_{0})\rfloor_{\bm{k}=\bm{k}_{\mathrm{FS}}}>0$.
    In Table 3, we plot contours of the FS with and without the order parameters which respect translational symmetry.  The FSs with non-zero order parameters are modified from the original FSs, but they are not fully gapped for all cases.

\subsection{Order parameters without translational symmetry}
Let us consider a spatially modulated order parameter which breaks a translational symmetry. We assume that an original band and its folded one after breaking translational symmetry are overlapped, generating hot-spots in the Brillouin zone. Depending on spin-rotational quantum numbers,  charge-density-wave (CDW) and spin-density-wave (SDW) are considered.

Our starting point is the two-band model. The hole (electron) pockets centered at $\Gamma$ ($M$) point are labeled by a quantum number $\nu=-(+)$ which specifies a dispersion relation, $E_{\nu}(\bm{k})=\epsilon_{0}(\bm{k})+\nu \sqrt{\epsilon_{3}(\bm{k})^{2}+\epsilon_{xy}(\bm{k})^{2}}$. The fluctuating order parameters carries a finite momentum $\bm{Q}$. We consider a specific momentum vector, $\bm{Q}\in \bm{Q}_{1},\bm{Q}_{2}$ which may form some hot spots ($\bm{Q}_{1}=(\pi,0)$, $\bm{Q}_{2}=(0,\pi)$). At hot spots, an original hole band and a shifted electron band intersect, $E_{-}(\bm{k})=E_{+}(\bm{k}+\bm{Q})=0$, and the corresponding order parameters are
\begin{eqnarray*}
\rho(\bm{r})=\rho_{\bm{Q}} e^{i\bm{Q}\cdot \bm{r}}&,&
\bm{M}(\bm{r})= \bm{M}_{\bm{Q}} e^{i\bm{Q}\cdot \bm{r}},
\end{eqnarray*}
with \vspace{-5pt}
\begin{eqnarray*}
\rho_{\bm{Q}}=\left\langle \sum_{\bm{k}}\psi_{\bm{k}}^{\dagger}\sigma_{0}\psi_{\bm{k}+\bm{Q}}\right\rangle&,&
\bm{M}_{\bm{Q}}=\left\langle \sum_{\bm{k}}\psi_{\bm{k}}^{\dagger}\bm{\sigma}\psi_{\bm{k}+\bm{Q}}\right\rangle.
\end{eqnarray*}

The dispersion relation is modified by the presence of the order parameters,
\begin{eqnarray}
E_{\mathrm{MF}}(\bm{k})&=&\frac{E_{-}(\bm{k})+E_{+}(\bm{k}+\bm{Q})}{2}\pm \left(\left(\frac{E_{-}(\bm{k})-E_{+}(\bm{k}+\bm{Q})}{2}\right)^{2}+\phi^{2}\right)^{1/2}.
\end{eqnarray}
where $\phi=\rho$ ($\phi=|\bm{M}|$) is for a CDW (SDW).
FSs with the order parameters are illustrated in Fig. S9. At hot spots, energy becomes are gapped, but a full gap opening is impossible near the onset of the order parameter.

Our analysis with and without translational symmetry may be straightforwardly generalized to include more bands. One crucial point is that a full gap opening may not appear without the onset of superconductivity. It is because superconductivity breaks $U(1)$ symmetry at all momentum points, so all the points may be considered as ``hot spots''.

\subsection{Estimation of gap anisotropy}
Now we focus on gap anisotropy associated with an order parameter which breaks a translational symmetry ($\bm{Q} \neq 0$).
The FSs at the $\Gamma$ point are characterized by $\bm{k}_{F}^{h}$ and $\bm{k}_{F}^{e}$, where
$E_{-}(\bm{k}_{F}^{h})=0$, $E_{+}(\bm{k}_{F}^{e}+\bm{Q}_{1})=0$.
The onset of an order parameter ($\phi \neq 0$) introduces a gap opening around the hot spots, which may be expressed as,
\begin{eqnarray}
\Delta^{h}(\theta_{h})&=& \Delta^{h}_{0}(\phi)+\Delta^{h}_{2}(\phi)\cos( 2\theta_{h}),\\
\Delta^{e}(\theta_{e})&=& \Delta^{e}_{0}(\phi)+\Delta^{e}_{2}(\phi)\cos (2\theta_{e}).
\end{eqnarray}
Polar angles of ($\bm{k}_{F}^{h},\bm{k}_{F}^{e}$) on each FS are introduced, ($\theta_{h},\theta_{e}$). We plot the ratios between a two-fold symmetric gap and an isotropic gap in Fig. S10.

\newpage
\begin{figure}[h]
	\centerline{\includegraphics[scale=1]{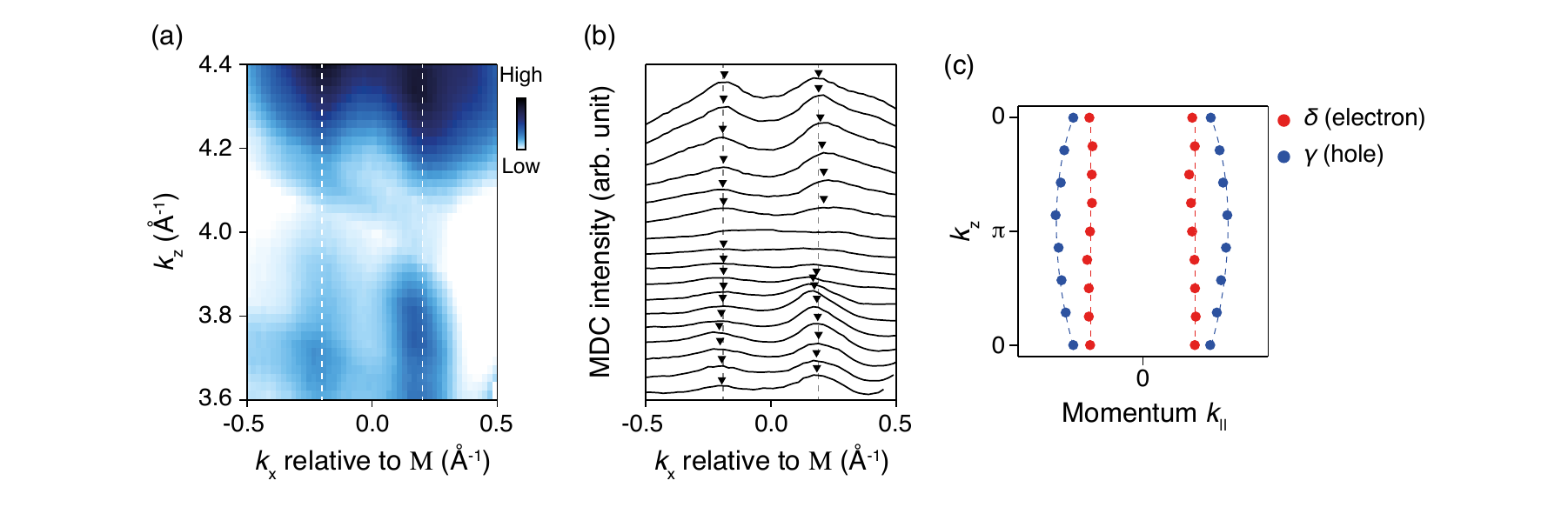}}
	\vspace{-5pt}
	\caption{(a) $k_z$-$k_x$ band dispersion of electron band near $M$ point, taken at $E_F$. Temperature was 30 K. White dashed lines are guide for eye. (b) MDC plots of (a) to see the dispersion more clearly. Black arrows indicate intensity maxima as guide for eye with grey dashed lines. (c) $k_z$ dependent peak positions of $\delta$ electron band (red) and $\gamma$ hole band (blue), taken from (a) and (b) ($\delta$), and Fig. 3d ($\gamma$), respectively. Red and blue dashed lines represent guide for eye. The center of $x$-axis corresponds to $k_x$ = $\pi$ for $\delta$ electron band and $k_x$ = 0 for $\gamma$ hole band.}
	\label{FigS1}
\end{figure}

\begin{figure}[h]
	\centerline{\includegraphics[scale=1]{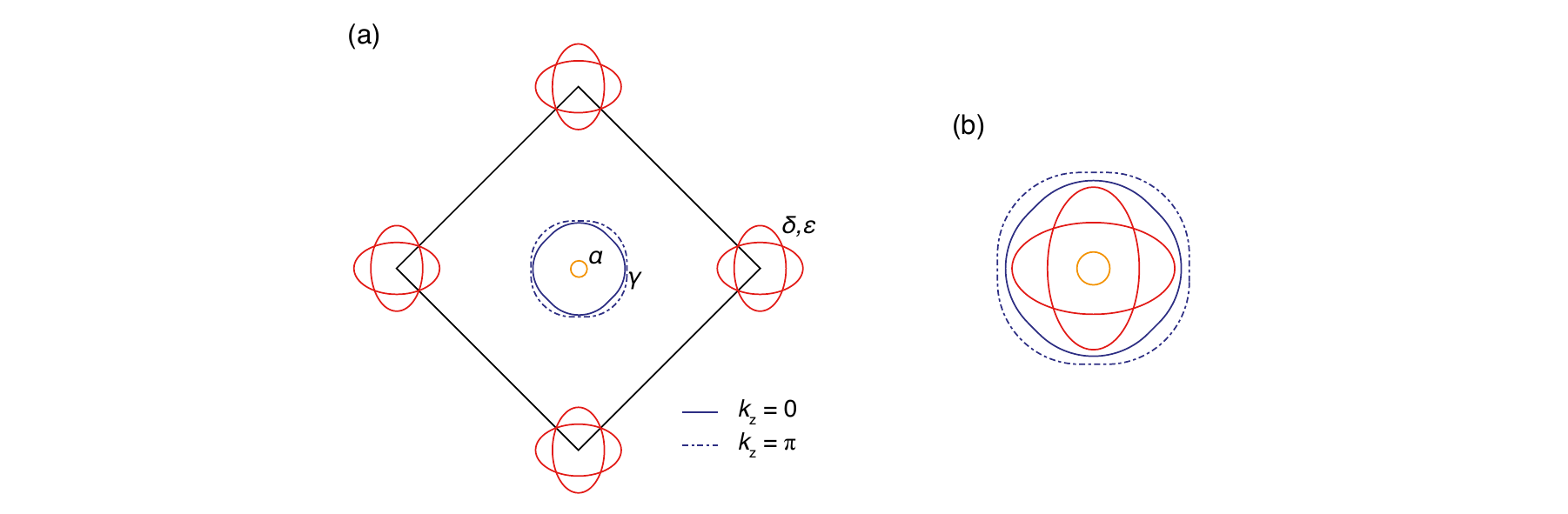}}
	\vspace{-5pt}
	\caption{(a) Schematic of FS of Sr$_2$VO$_3$FeAs. Black solid line indicates 2 Fe unit cell BZ. Blue solid and dashed lines represent $\gamma$ hole pocket in $k_z$ = 0 and $k_z$ = $\pi$ plane, respectively. The size of pockets are extracted from the experimental data. (b) All electron and hole pockets are overlapped to compare their FS size.}
	\label{FigS2}
\end{figure}
\newpage
\begin{figure}[h]
	\centerline{\includegraphics[scale=1]{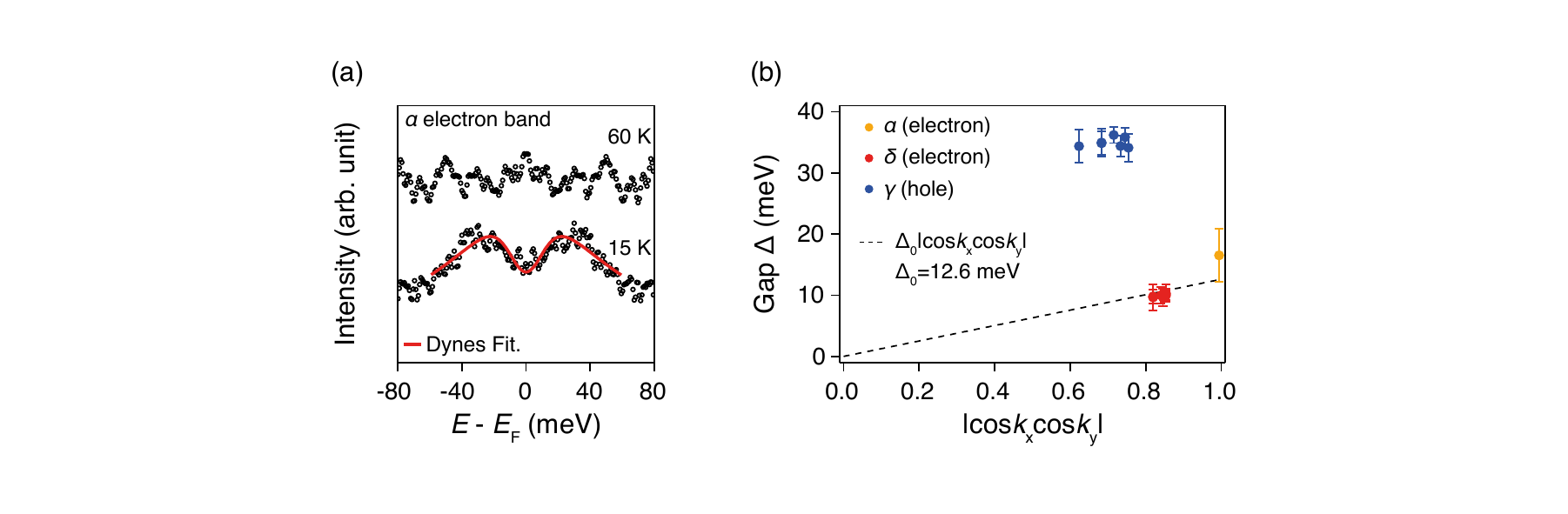}}
	\vspace{-5pt}
	\caption{(a) Symmetrized EDCs taken on $\alpha$ electron band at two different temperature of 15 K and 60 K, which corresponds to below and above $T_C$. Red solid curve shows fit result of 15 K spectrum with Dynes function. At low temperature, peak and dip feature of superconducting gap is clearly observed. (b) Plot of gap size as function of $|\cos{k_x}\cos{k_y}|$. Black dashed line exhibit a fitting result assuming the gap function $\Delta$ = $\Delta_0$ $|\cos{k_x}\cos{k_y}|$, excluding $\gamma$ hole band. Estimated $\Delta_0$ is 12.6 meV.}
	\label{FigS3}
\end{figure}

\begin{figure}[h]
	\centerline{\includegraphics[scale=1]{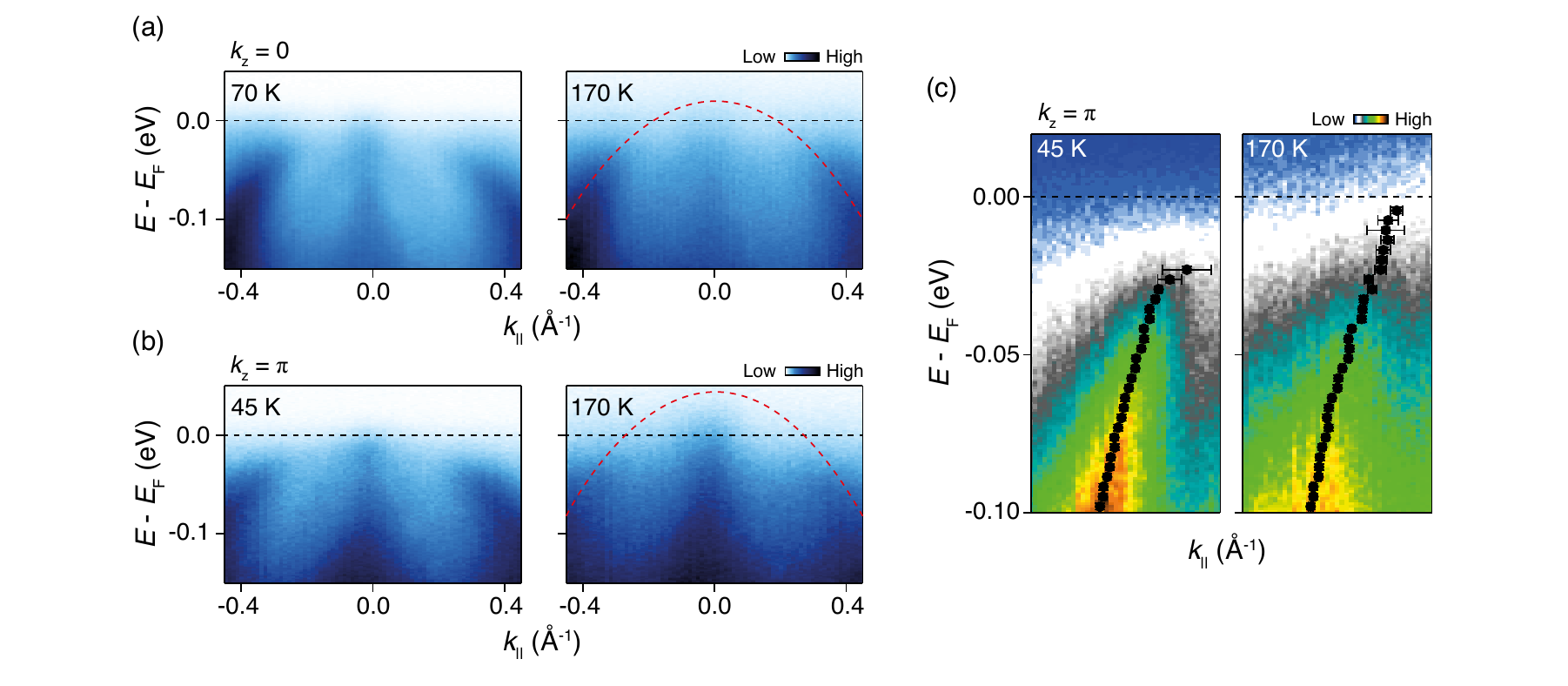}}
	\vspace{-5pt}
	\caption{(a),(b) ARPES results along the (0,0)-($\pi$, $\pi$) direction at $k_z$ = 0 (a) and $k_z$ = $\pi$ (b) plane, respectively. Left and right panels represent the results observed at below and above $T_{\rm HO}$ (150 K), respectively. Red dashed lines on right panels are guides for eye to trace hole band dispersion. (c) Expanded view of the $\gamma$ hole band from (b). Peak positions determined from MDCs are overlaid with black filled circles.}
	\label{FigS4}
\end{figure}
\newpage
\begin{figure}[h]
	\centerline{\includegraphics[scale=1]{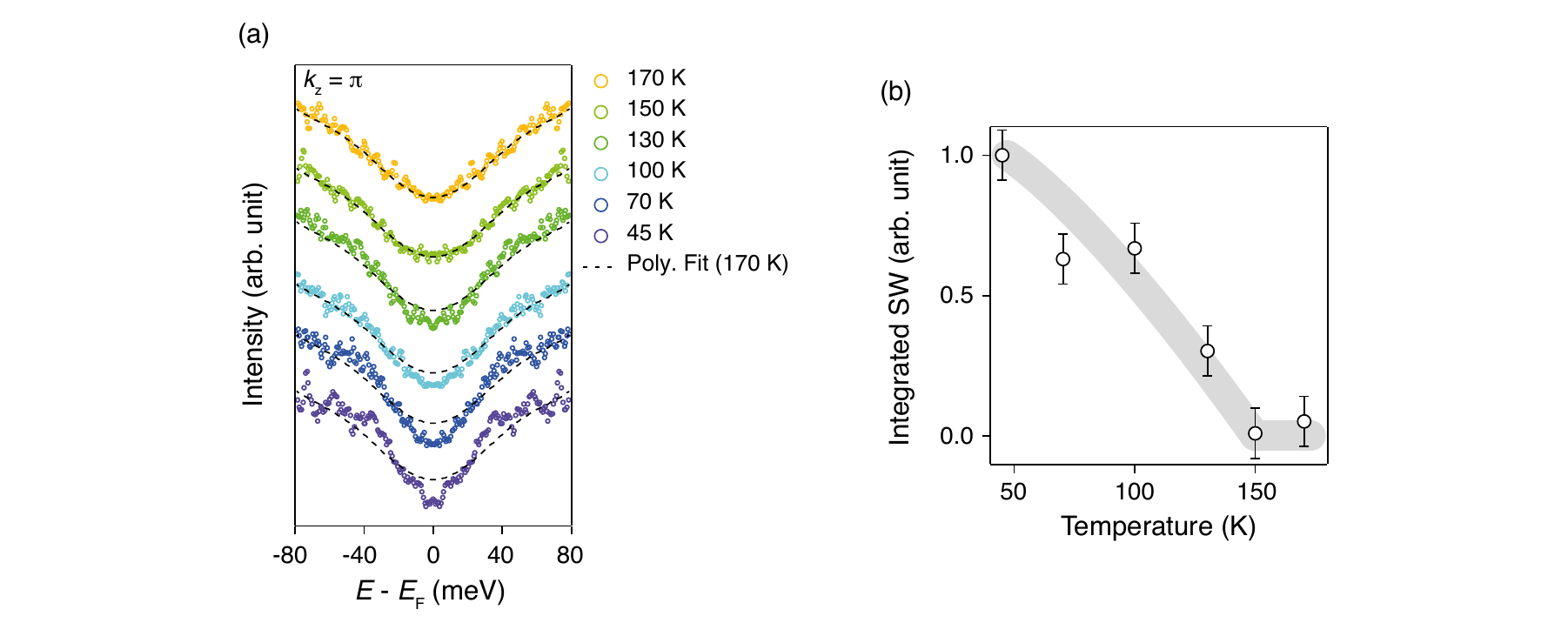}}
	\vspace{-5pt}
	\caption{(a) Symmetrized EDCs from the $\gamma$ hole band in $k_z$ = $\pi$ plane at temperatures between 45 and 170 K. Overlaid black dashed lines are the polynomial fit curves of the 170 K spectrum. (b) Integrated spectral weight difference with respect to polynomial fit curve of the 170 K spectrum as a function of the temperature. The integrated spectral weight of 45 K is normalized to 1.}
	\label{FigS5}
\end{figure}

\begin{figure}[h]
	\centerline{\includegraphics[scale=1]{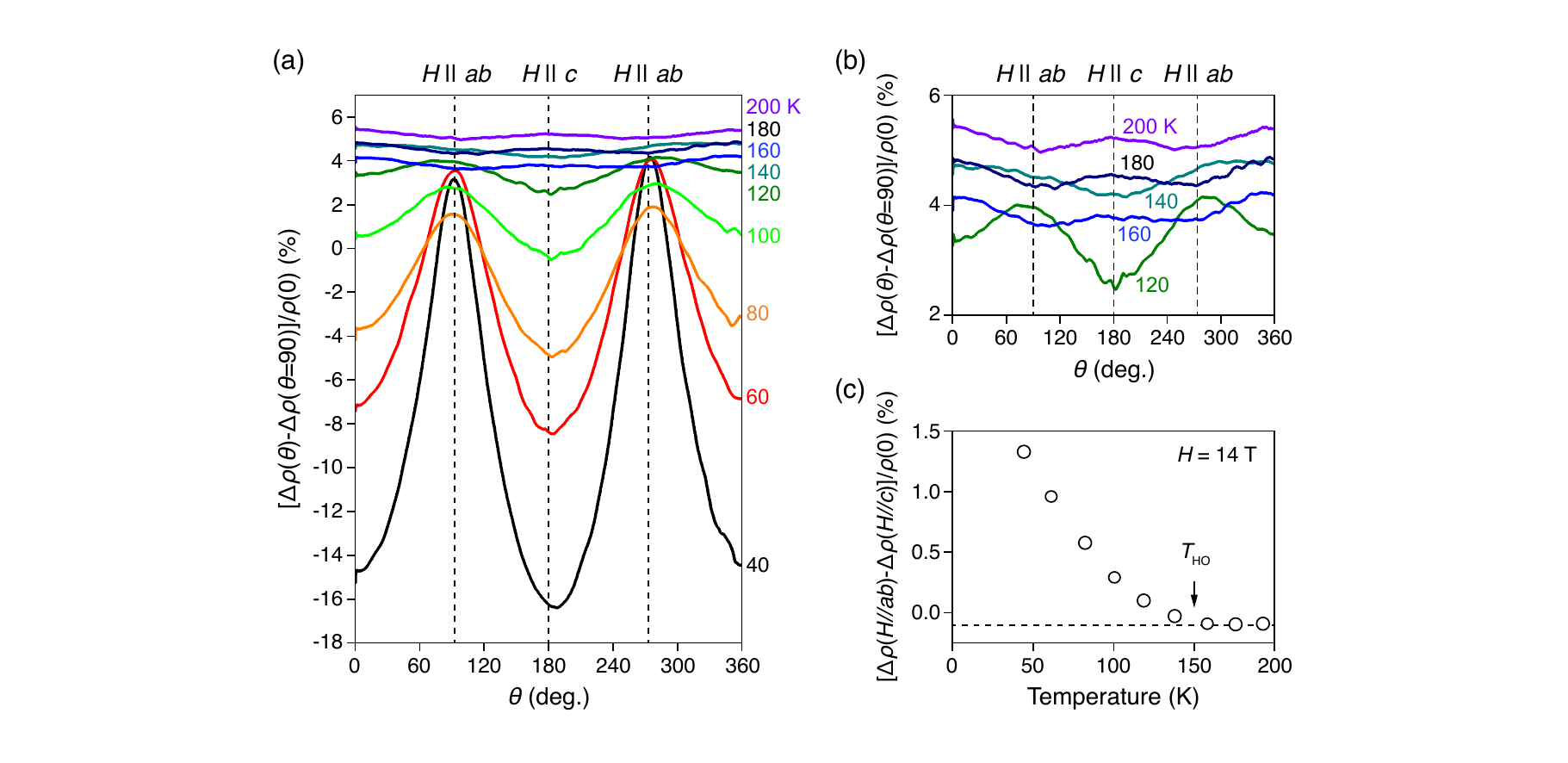}}
	\vspace{-5pt}
	\caption{(a),(b) Angle dependence of the MR at various temperatures. The opposite angle dependence across $T_{\rm HO}$ is clearly shown in (b). (c) The difference between $\Delta\rho (H\parallel ab)$ and $\Delta \rho (H\parallel c)$, normalized by $\rho(0)$. The onset of anisotropy of the MR well matches with $T_{\rm HO}$.}
	\label{FigS6}
\end{figure}
\newpage
\begin{figure}[h]
	\centerline{\includegraphics[scale=1]{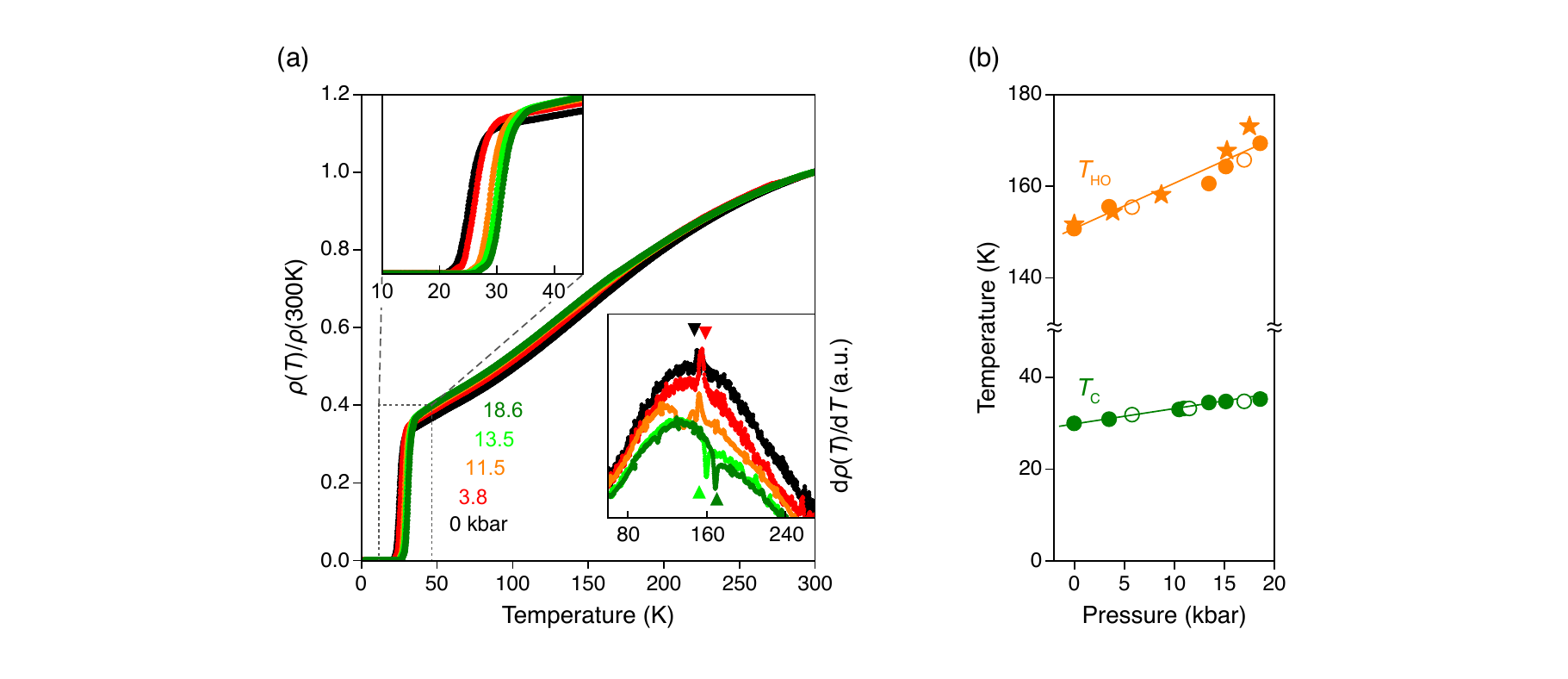}}
	\vspace{-5pt}
	\caption{(a) Temperature dependent in-plane resistivity ($\rho(T)$) at different pressures. The inset (left top) shows the resistive superconducting transitions, shifting systematically to higher temperatures with pressure. The temperature-derivative curves of $d\rho(T)/dT$ for different pressures are also presented in the inset (right bottom). The hidden order transition temperature $T_{\rm HO}$ is estimated by the anomalies in $d\rho(T)/dT$, indicated by triangles in the inset. (b) Pressure dependence of $T_C$ and $T_{\rm HO}$, taken from two samples, S1 (circle) and S2 (star). The $T_{\rm HO}$ data during increasing (solid) and decreasing (open) pressures show the consistent pressure dependence. The solid lines are the guide for the eye.}
	\label{FigS7}
\end{figure}
\newpage
\begin{figure}[h]
	\centerline{\includegraphics[scale=1]{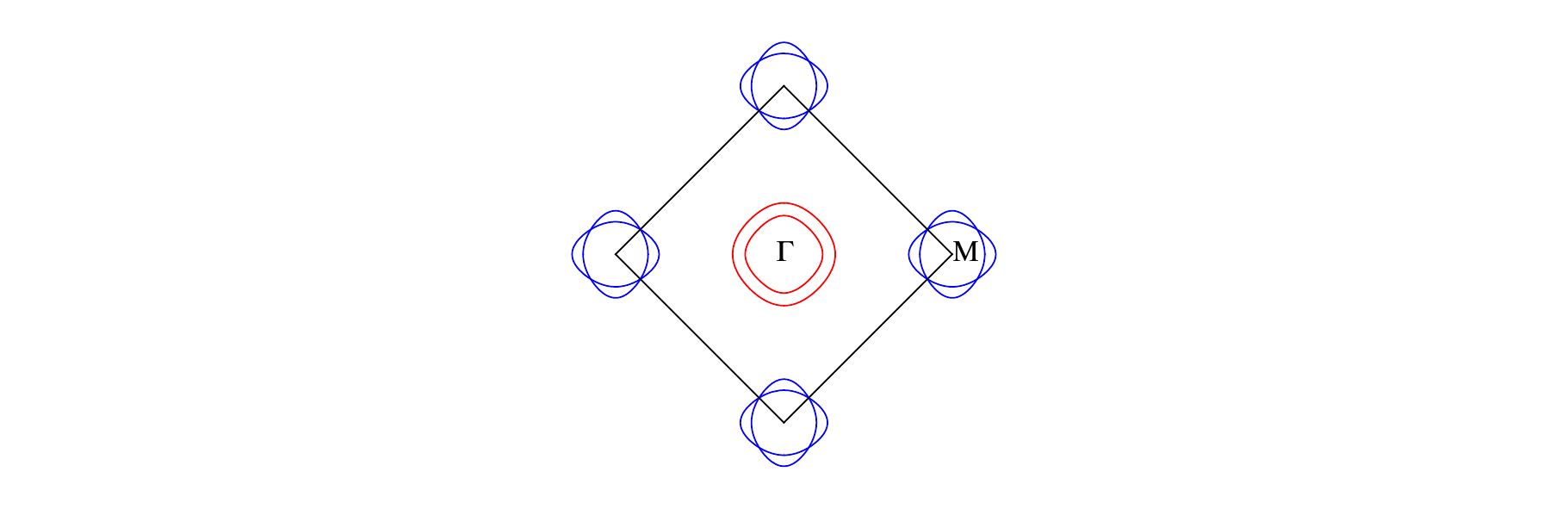}}
	\vspace{-5pt}
	\caption{A FS of $H_0$ with  $t_{1}=-1, t_{2}=1.3, t_{3}=t_{4}=-0.85$. The blue (red) line is for two hole (electron) pockets at $\Gamma$ ($M$).  Each band is doubly degenerate with spins.}
	\label{FigS8}
\end{figure}

\begin{figure}[h]
	\centerline{\includegraphics[scale=1]{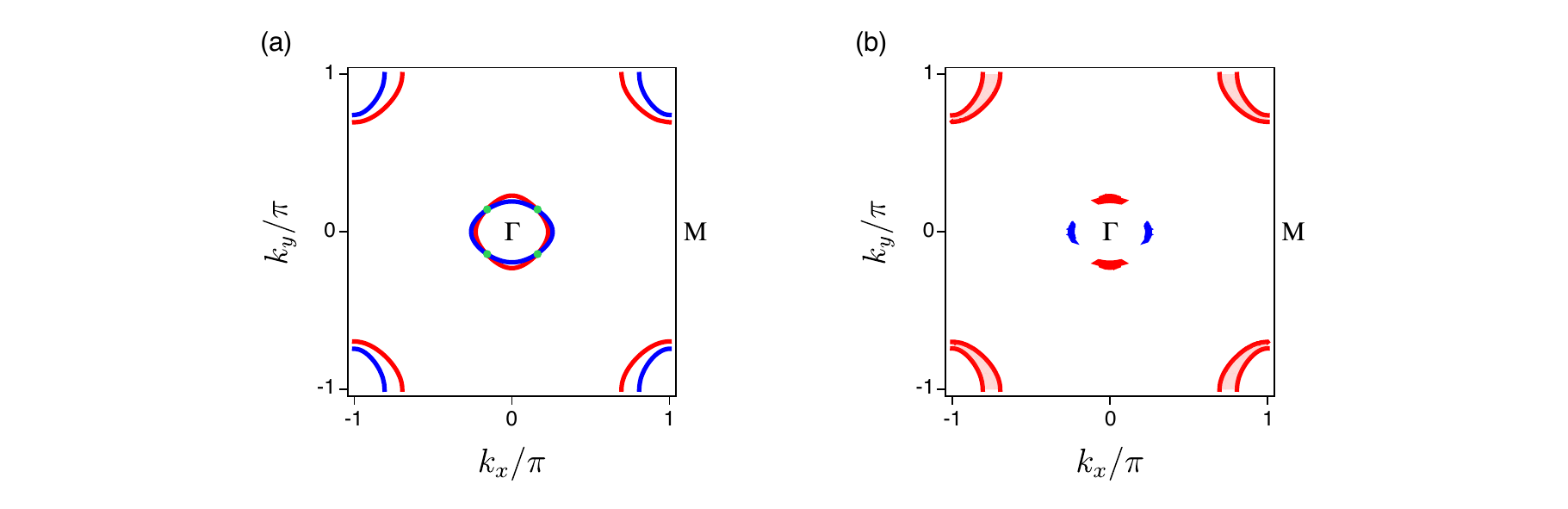}}
	\vspace{-5pt} 
	\caption{(a) Original FSs of hole pocket  ($E_{-}(\bm{k})$=0) and ones of electron pocket ($E_{+}(\bm{k}+\bm{Q}_1)$=0) in the unfolded BZ. The green points near $\Gamma$ are hot spots. (b) Modified FSs with the onset of an order parameter, $\rho$ or $\bm{M}$.}
	\label{FigS9}
\end{figure}

\begin{figure}[h]
	\centerline{\includegraphics[scale=1]{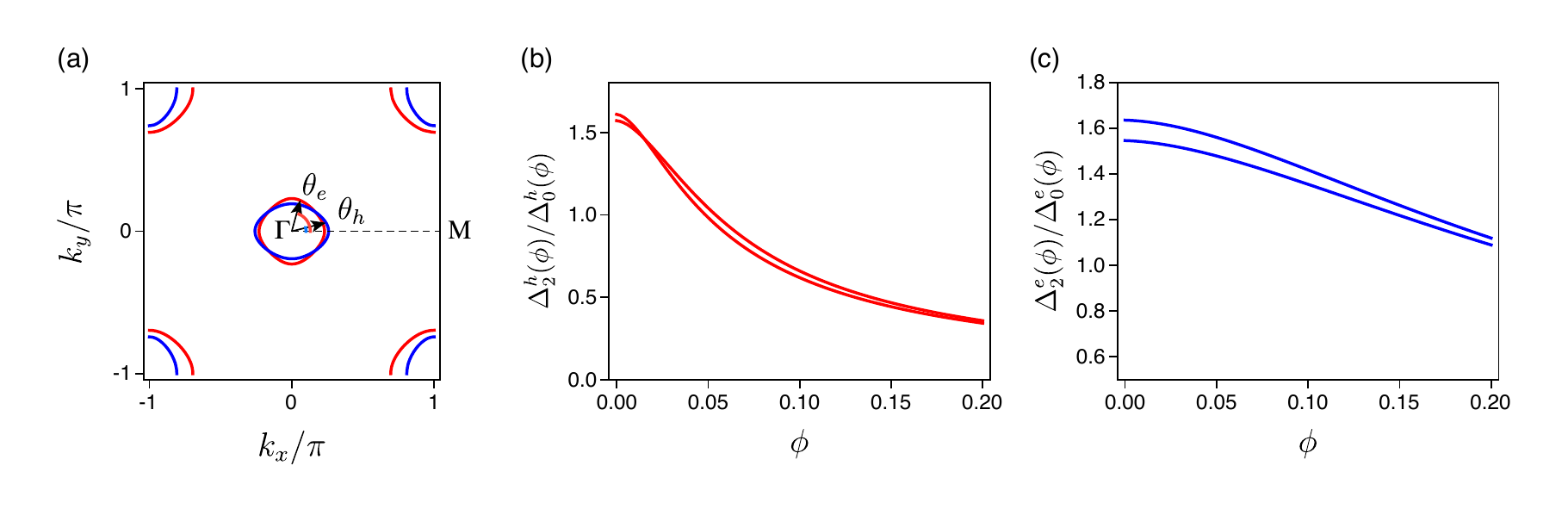}}
	\caption{(a) Original FSs of hole pocket  ($E_{-}(\bm{k})$=0) and ones of electron pocket ($E_{+}(\bm{k}+\bm{Q}_1)$=0) in the unfolded BZ. (b),(c) The gap ratios, $\Delta_{2}/\Delta_{0}$ at $\bm{k}=\bm{k}_{F}^{h}$, $\bm{k}=\bm{k}_{F}^{e}$, respectively, as a function of $\phi$. At $\phi\simeq 0.72$, all FSs are gapped in the BZ.}
	\label{FigS10}
\end{figure}
\newpage

\begin{table}[h]
\begin{center}
\renewcommand{\arraystretch}{1.3}
\begin{tabular}{C{1.2 cm}C{1.2cm}C{1.2cm}}
\hline
\hline
$\tau^{a}$ & $ \Gamma$ & $\mathcal{T} $\tabularnewline
\hline
$\tau^{0}$ &$A_{1}$ &$+1$  \tabularnewline
$\tau^{1}$ &$B_{2}$ &$+1$ \tabularnewline
$\tau^{2}$ &$A_{2}$ &$-1$ \tabularnewline
$\tau^{3}$ &$B_{1}$ &$+1$ \tabularnewline
\hline
\hline
\end{tabular}\ \ \ \ \
\begin{tabular}{C{1.3 cm}C{1.8cm}C{1.2cm}}
\hline
\hline
$\sigma^{a}$ & Spin-Rot. & $\mathcal{T} $\tabularnewline
\hline
$\sigma^{0}$ &Yes&$+1$ \tabularnewline
$\vec{\sigma}$ &No &$-1$ \tabularnewline
\hline
\hline
\end{tabular}\caption{Symmetry operations of an orbital ($\tau^{a}$) and a spin ($\sigma^{a}$) part. }\label{t1}
\end{center}
\end{table}

\begin{table}[h]
\begin{center}
\renewcommand{\arraystretch}{1.3}
\begin{tabular}{	C{1.2 cm} C{2.5cm}C{4cm}C{3.6cm}}
\hline
\hline
$\Gamma$ &$\psi_{\Gamma}(\bm{k})$ &$M^{\Gamma,+}_{o}(\bm{k})$, $M^{\Gamma,-}_{s}(\bm{k})$&$M^{\Gamma,-}_{o}(\bm{k})$, $M^{\Gamma,+}_{s}(\bm{k})$ \tabularnewline
\hline
$A_{1}$ &$k_{x}^{2}+k_{y}^{2}$ & $\psi_{A_{1}}\tau^{0},\psi_{B_{1}}\tau^{3},\psi_{B_{2}}\tau^{1}$&$\psi_{A_{2}}\tau^{2}$ \tabularnewline
$A_{2}$ &$k_{x}k_{y}(k_{x}^{2}-k_{y}^{2})$& $\psi_{A_{2}}\tau^{0},\psi_{B_{2}}\tau^{3},\psi_{B_{1}}\tau^{1}$&$\psi_{A_{1}}\tau^{2}$ \tabularnewline
$B_{1}$ &$k_{x}^{2}-k_{y}^{2}$& $\psi_{B_{1}}\tau^{0},\psi_{A_{1}}\tau^{3},\psi_{A_{2}}\tau^{1}$&$\psi_{B_{2}}\tau^{2}$ \tabularnewline
$B_{2}$ &$k_{x}k_{y}$& $\psi_{B_{2}}\tau^{0},\psi_{A_{2}}\tau^{3},\psi_{A_{1}}\tau^{1}$&$\psi_{B_{1}}\tau^{2}$ \tabularnewline
$E$&$(k_{x},k_{y})$ & $\vec{\psi}_{E}\tau^{0},\ \vec{\psi}_{E}\tau^{3},\ \vec{\psi}_{E}\tau^{1}$&$\vec{\psi}_{E}\tau^{2}$ \tabularnewline
\hline
\hline
\end{tabular} \vspace{5pt}
\caption{ The basis functions of a point group $D_{4}$ (the second column). The matrix forms for order parameters $\phi^{\Gamma,\mathcal{T}}_{s,o}$ (the third and fourth column). The superscript ($\Gamma,\mathcal{T}$) is for the symmetry action under the point group and time-reversal symmetry. The subscript ($s,o$) is for the spin singlet and triplet channels.  }
\label{t2}
\end{center}
\end{table}

\begin{table}[h]
\begin{center}
\renewcommand{\arraystretch}{1.5}
\begin{tabular}{C{.8cm}C{2.8cm}C{2.8cm}}
\hline
\hline
&$\mathcal{T}=1$&$\mathcal{T}=-1$\\
\hline
\rule{0pt}{40pt}\begin{minipage}{0\textwidth}\begin{center}
 $\phi^{A_{1},\mathcal{T}}_{o}$
\end{center}
\end{minipage}  &
\begin{minipage}{0.2\textwidth}\begin{center}
\includegraphics[scale=0.18]{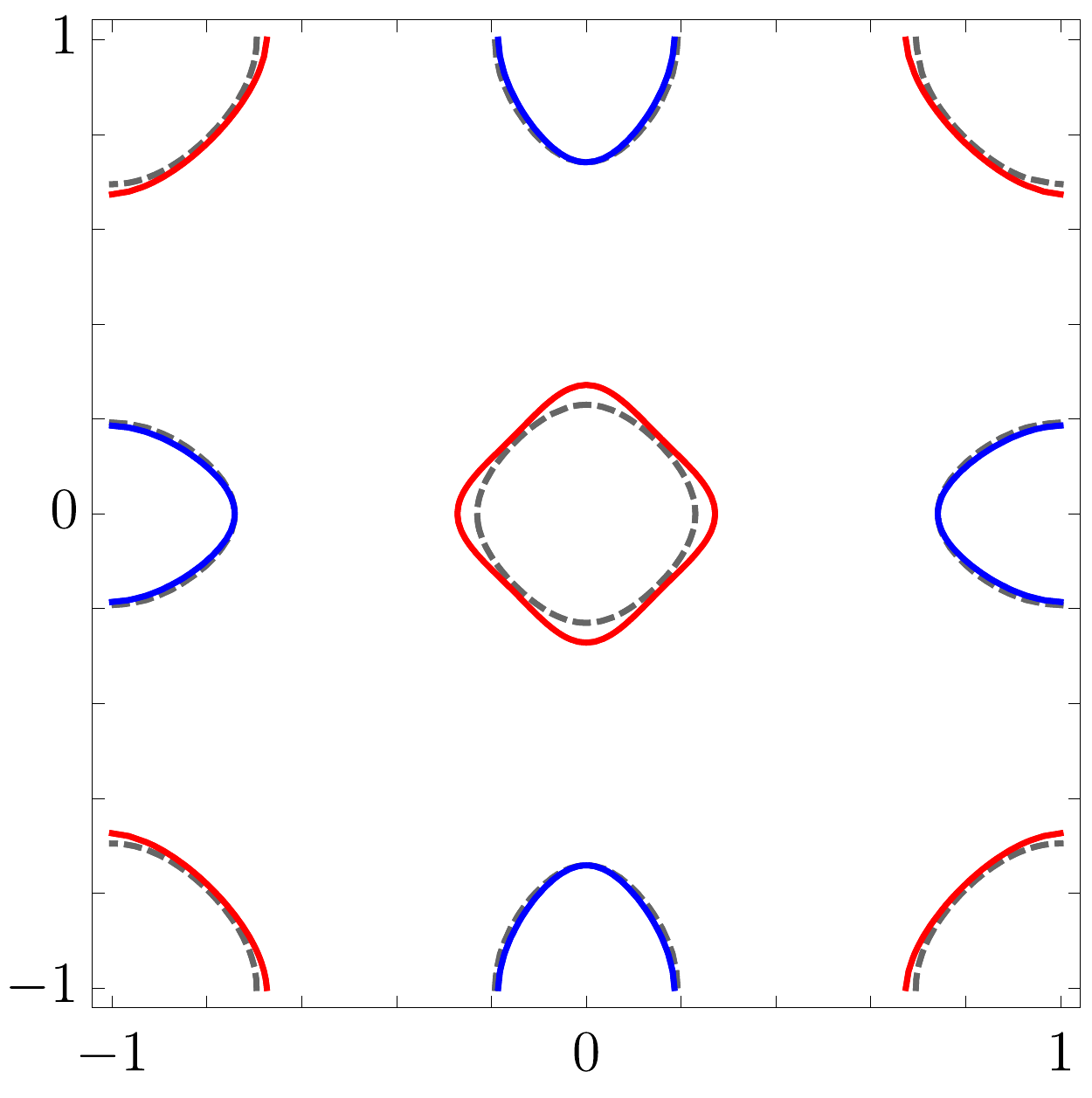}
\end{center}
\end{minipage}
&\begin{minipage}{0.2\textwidth}
\begin{center}
\includegraphics[scale=0.18]{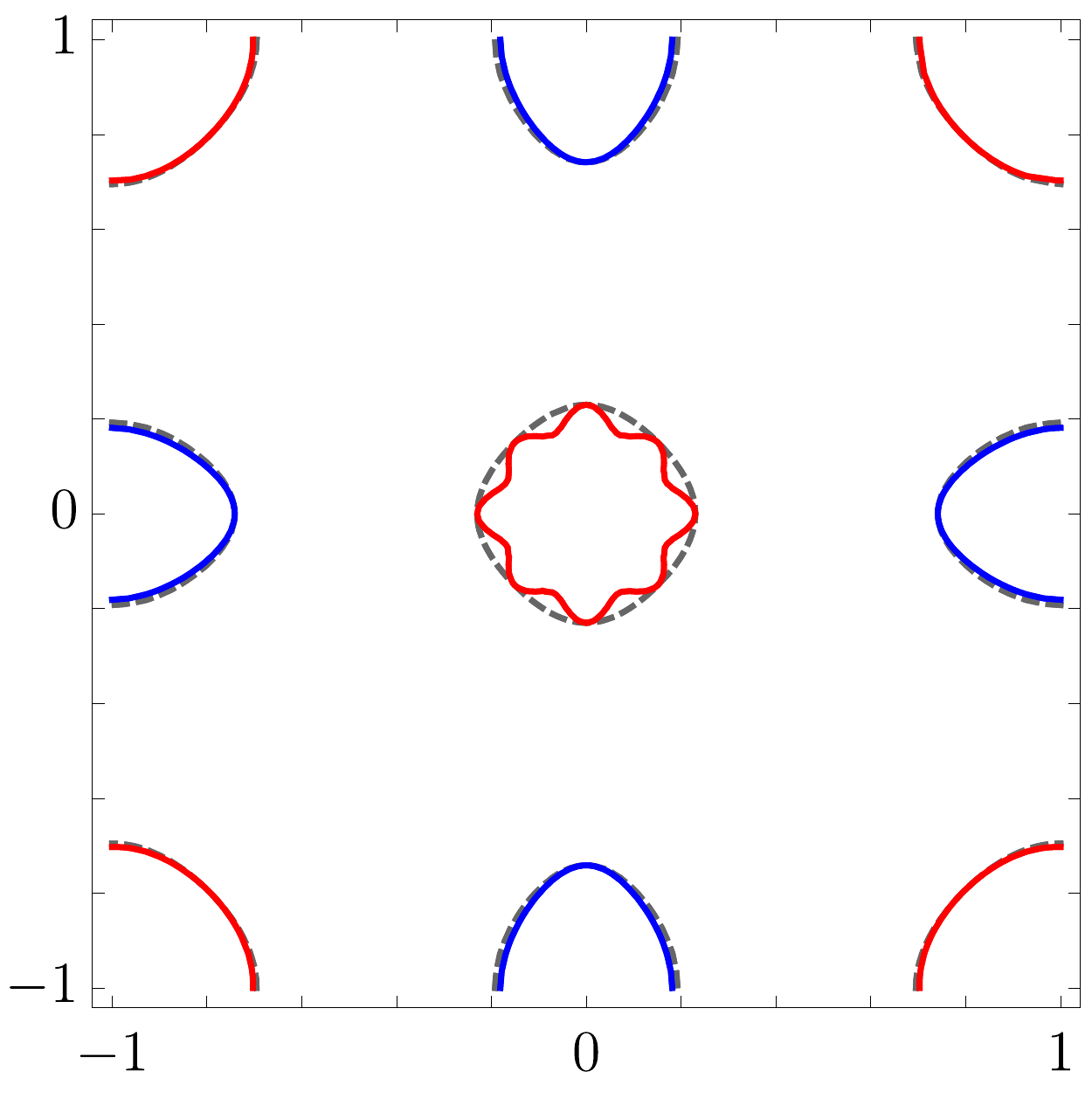}
\end{center}
\end{minipage}
\\
\rule{0pt}{40pt}\rule{0pt}{40pt}\begin{minipage}{0\textwidth}\begin{center}
 $\phi^{A_{2},\mathcal{T}}_{o}$
\end{center}
\end{minipage}   &
\begin{minipage}{0.2\textwidth}
\begin{center}
\includegraphics[scale=0.18]{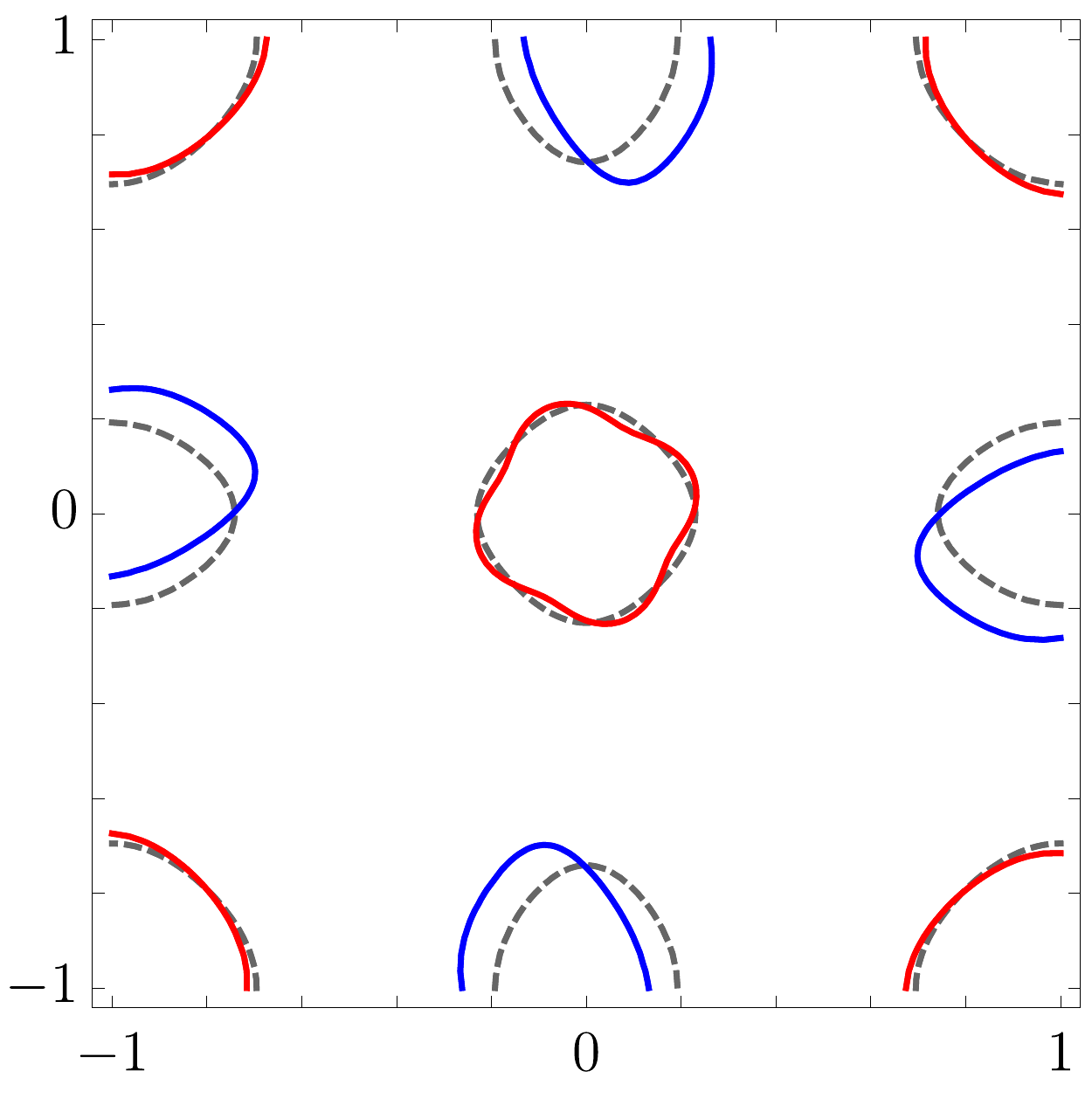}
\end{center}
\end{minipage}
&\begin{minipage}{0.2\textwidth}
\begin{center}
\includegraphics[scale=0.18]{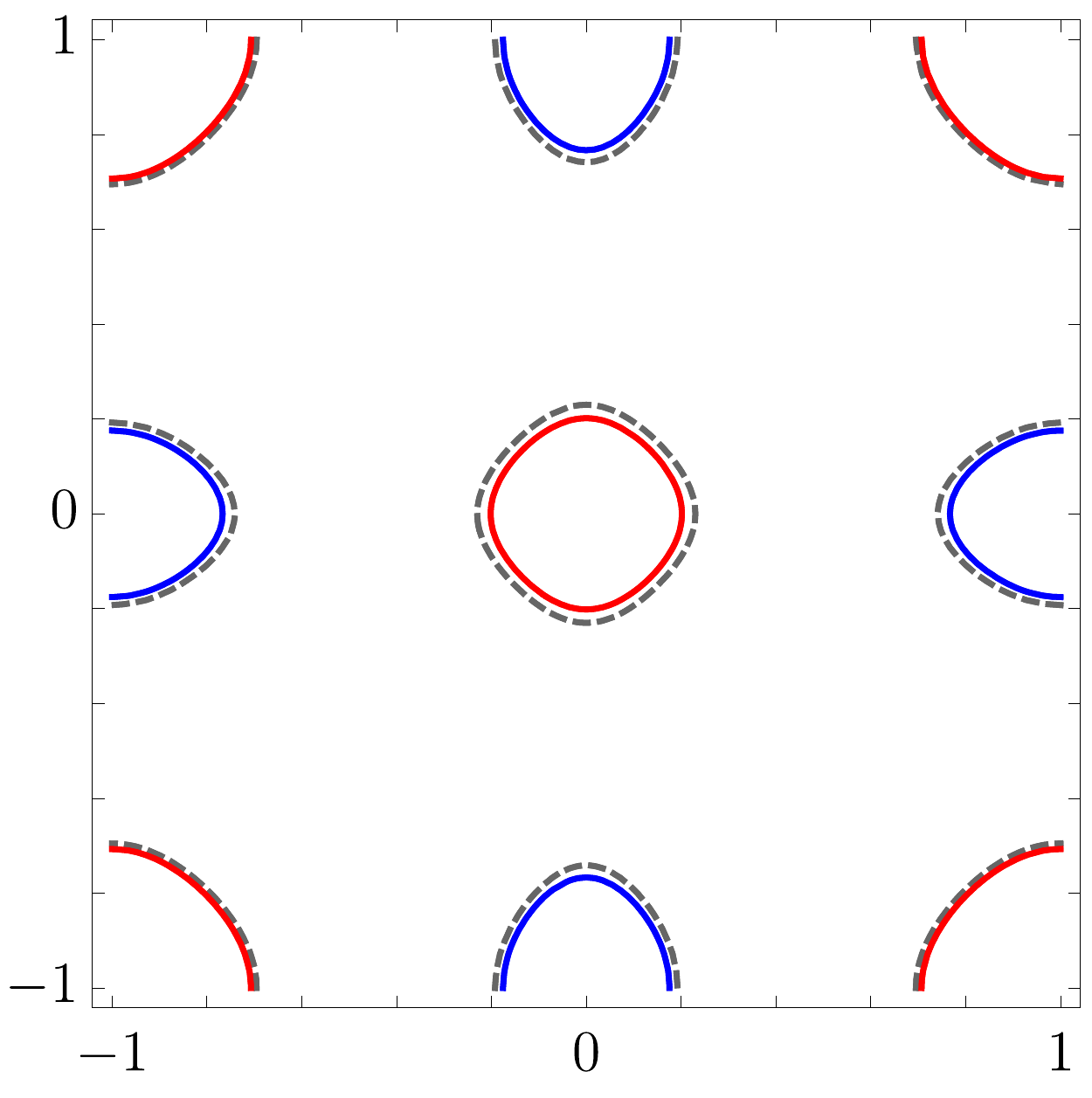}
\end{center}
\end{minipage}\\
\rule{0pt}{40pt} \begin{minipage}{0\textwidth}\begin{center}
 $\phi^{B_{1},\mathcal{T}}_{o}$
\end{center}
\end{minipage}   &
\begin{minipage}{0.2\textwidth}\begin{center}
\includegraphics[scale=0.18]{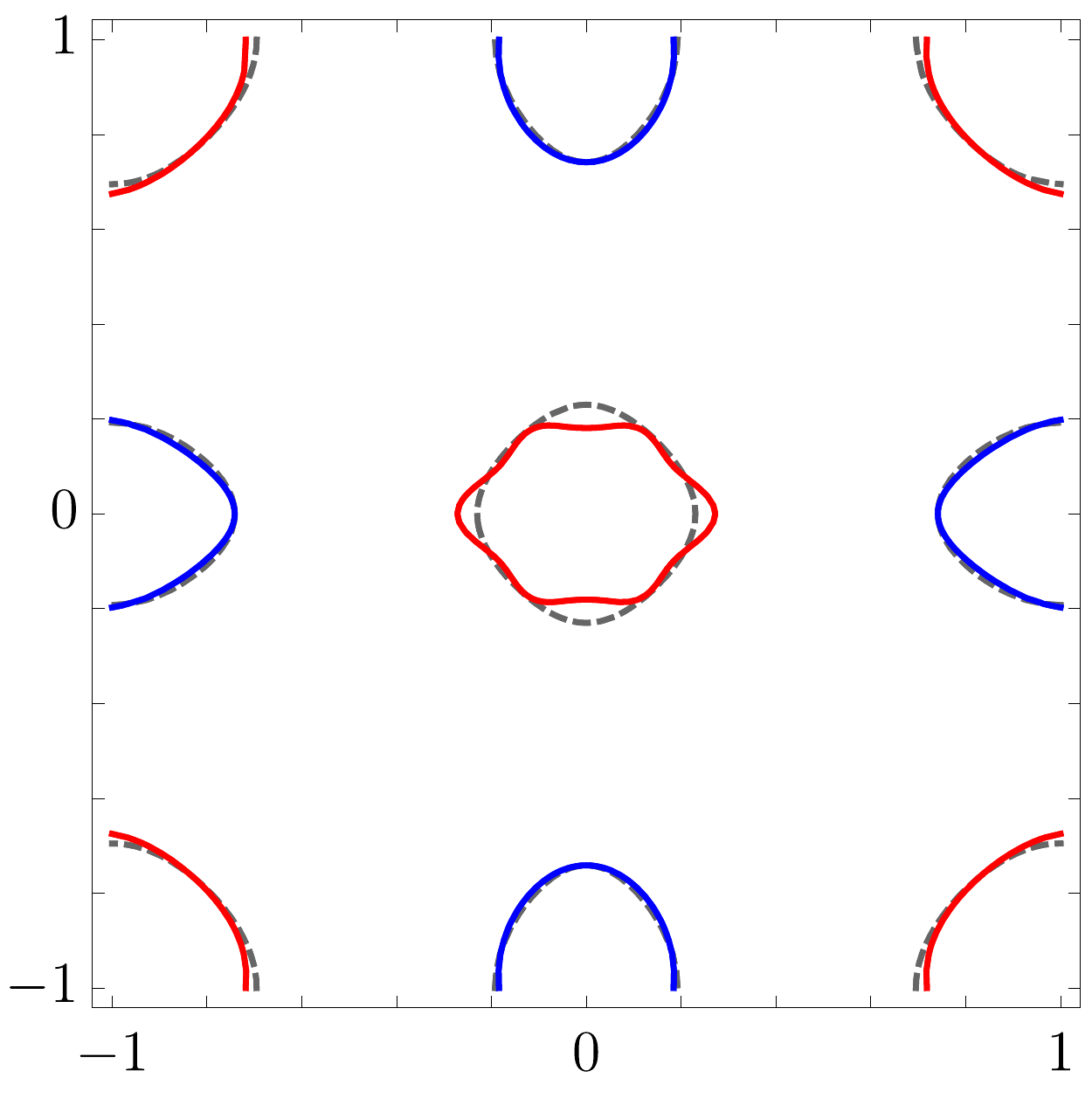}
\end{center}
\end{minipage}
& \begin{minipage}{0.2\textwidth}
\begin{center}
\includegraphics[scale=0.18]{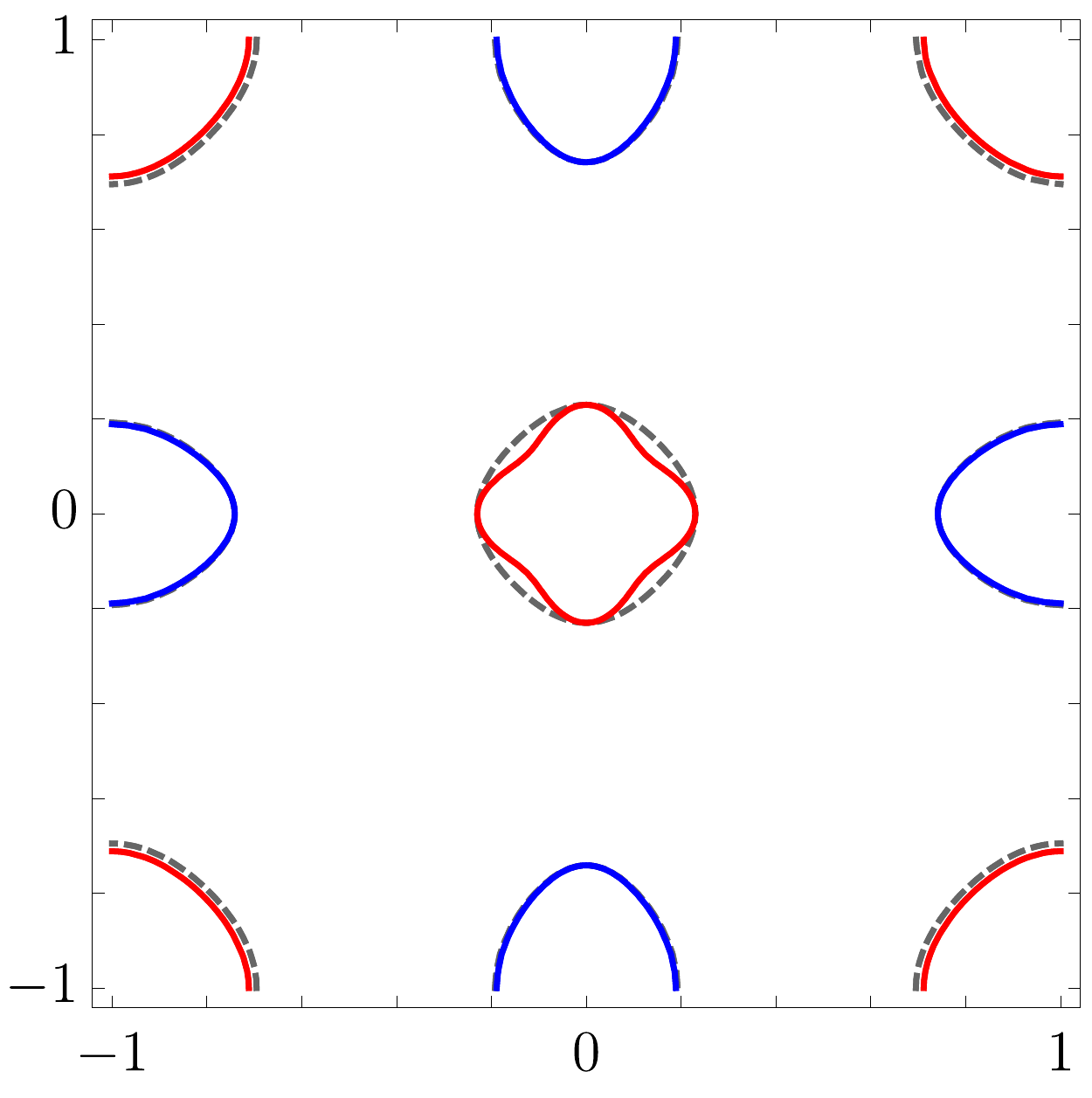}
\end{center}
\end{minipage}\\
\rule{0pt}{40pt}\begin{minipage}{0\textwidth}\begin{center}
 $\phi^{B_{2},\mathcal{T}}_{o}$
\end{center}
\end{minipage}  &
\begin{minipage}{0.2\textwidth}\begin{center}
\includegraphics[scale=0.18]{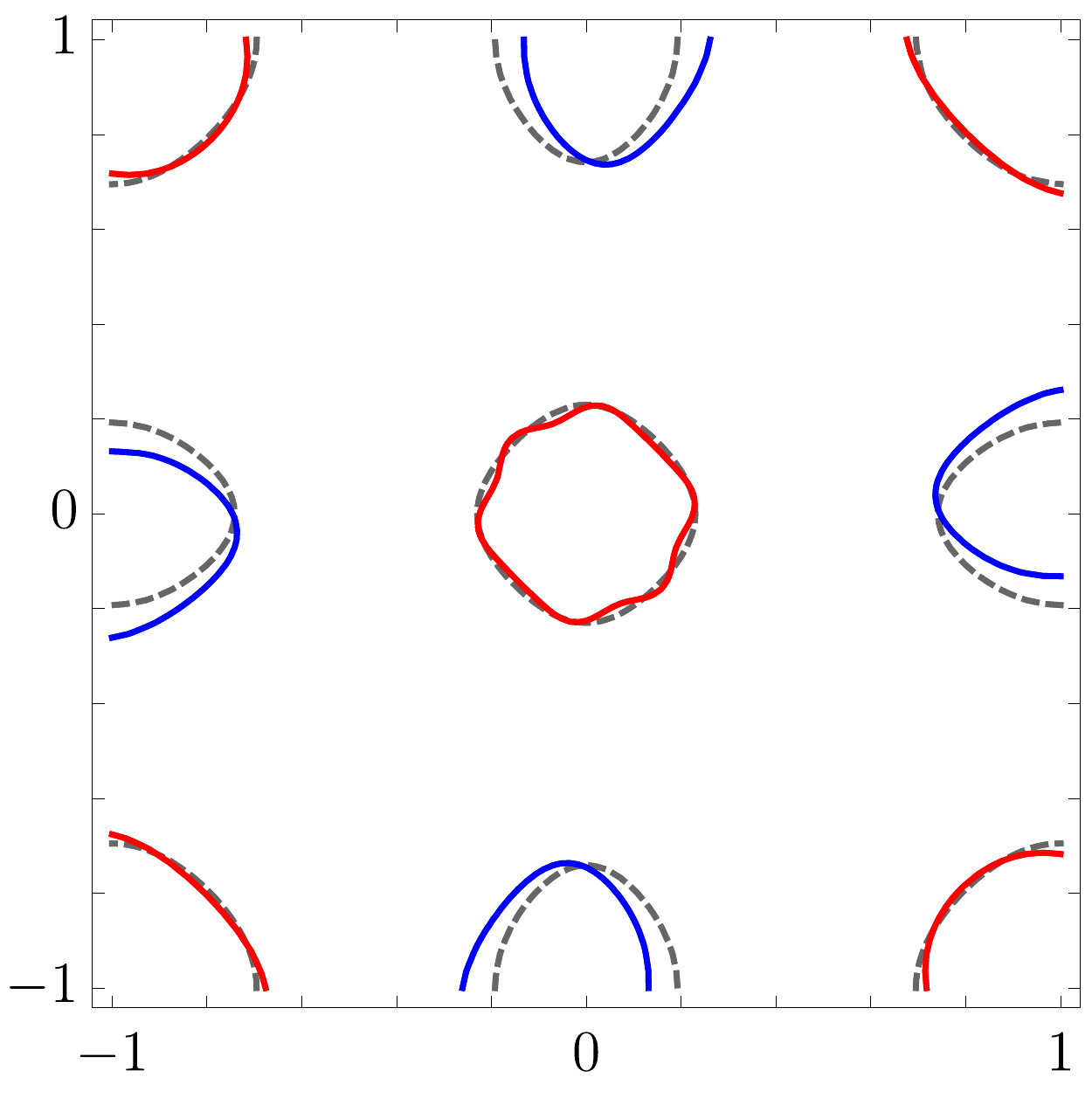}
\end{center}
\end{minipage}
& \begin{minipage}{0.2\textwidth}
\begin{center}
\includegraphics[scale=0.18]{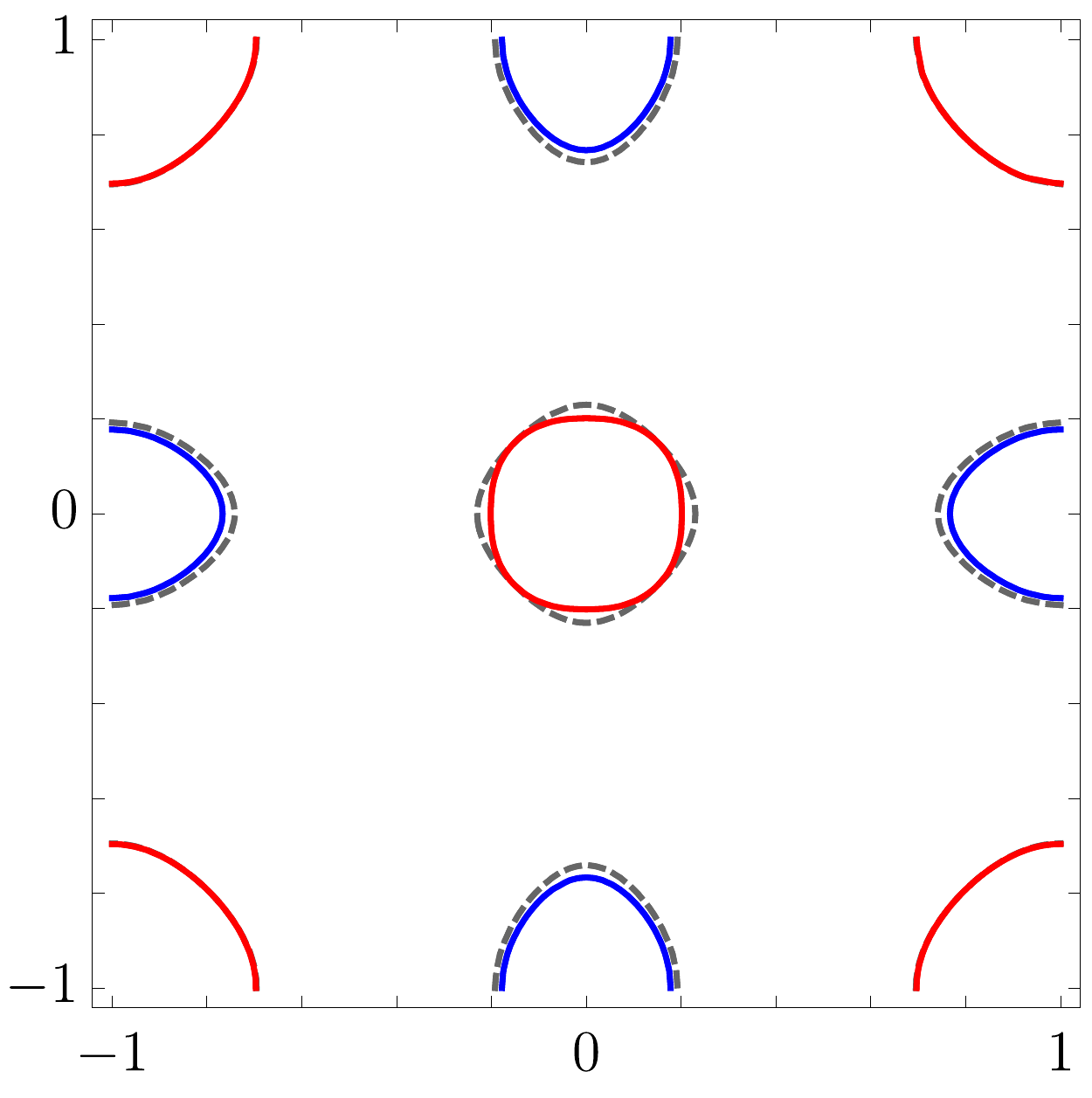}
\end{center}
\end{minipage}\\
\rule{0pt}{40pt} \begin{minipage}{0\textwidth}\begin{center}
 $\phi^{E,\mathcal{T}}_{o}$
\end{center}
\end{minipage}  &
\begin{minipage}{0.2\textwidth}\begin{center}
\includegraphics[scale=0.18]{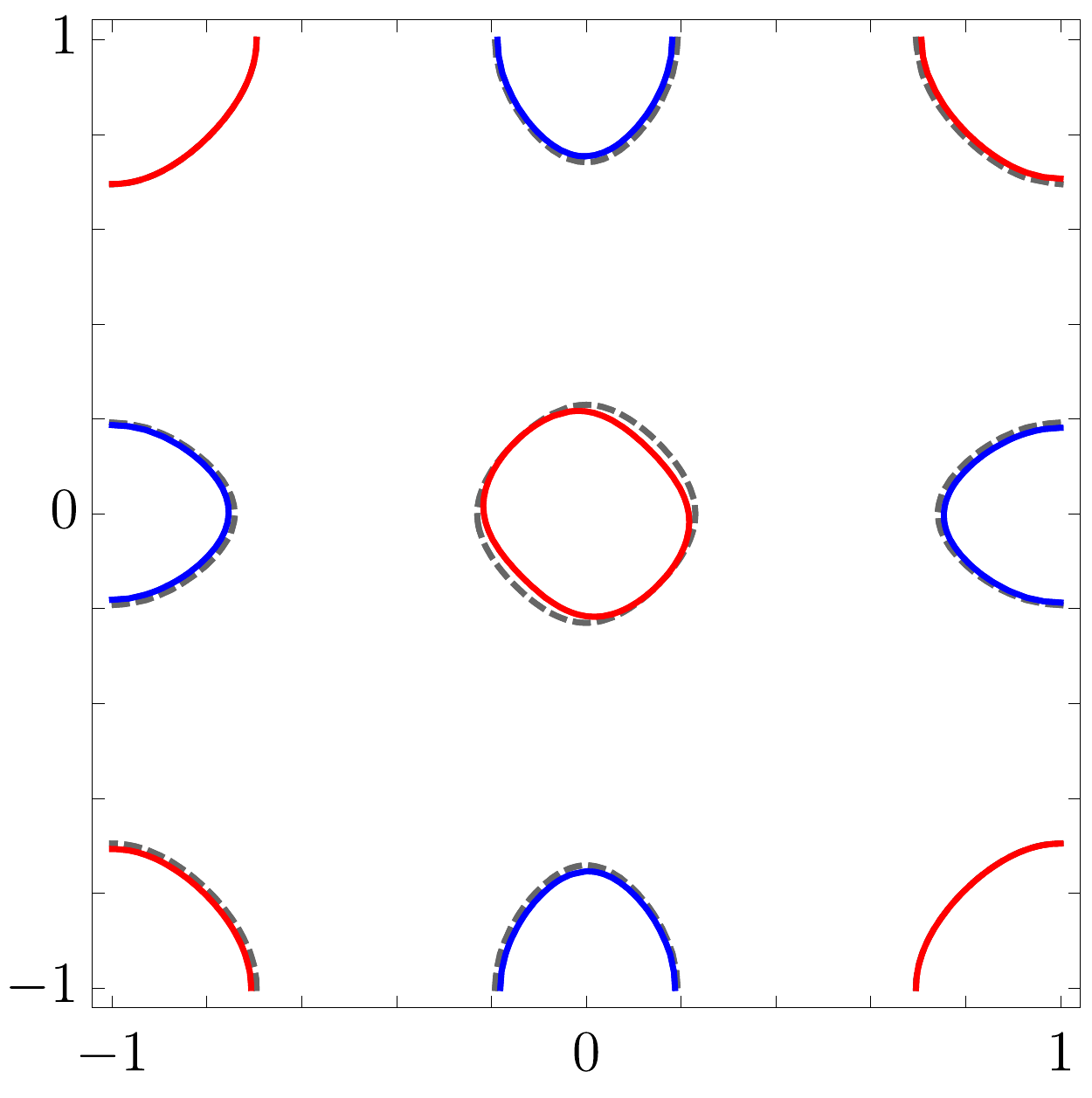}
\end{center}
\end{minipage}
& \begin{minipage}{0.2\textwidth}
\begin{center}
\includegraphics[scale=0.18]{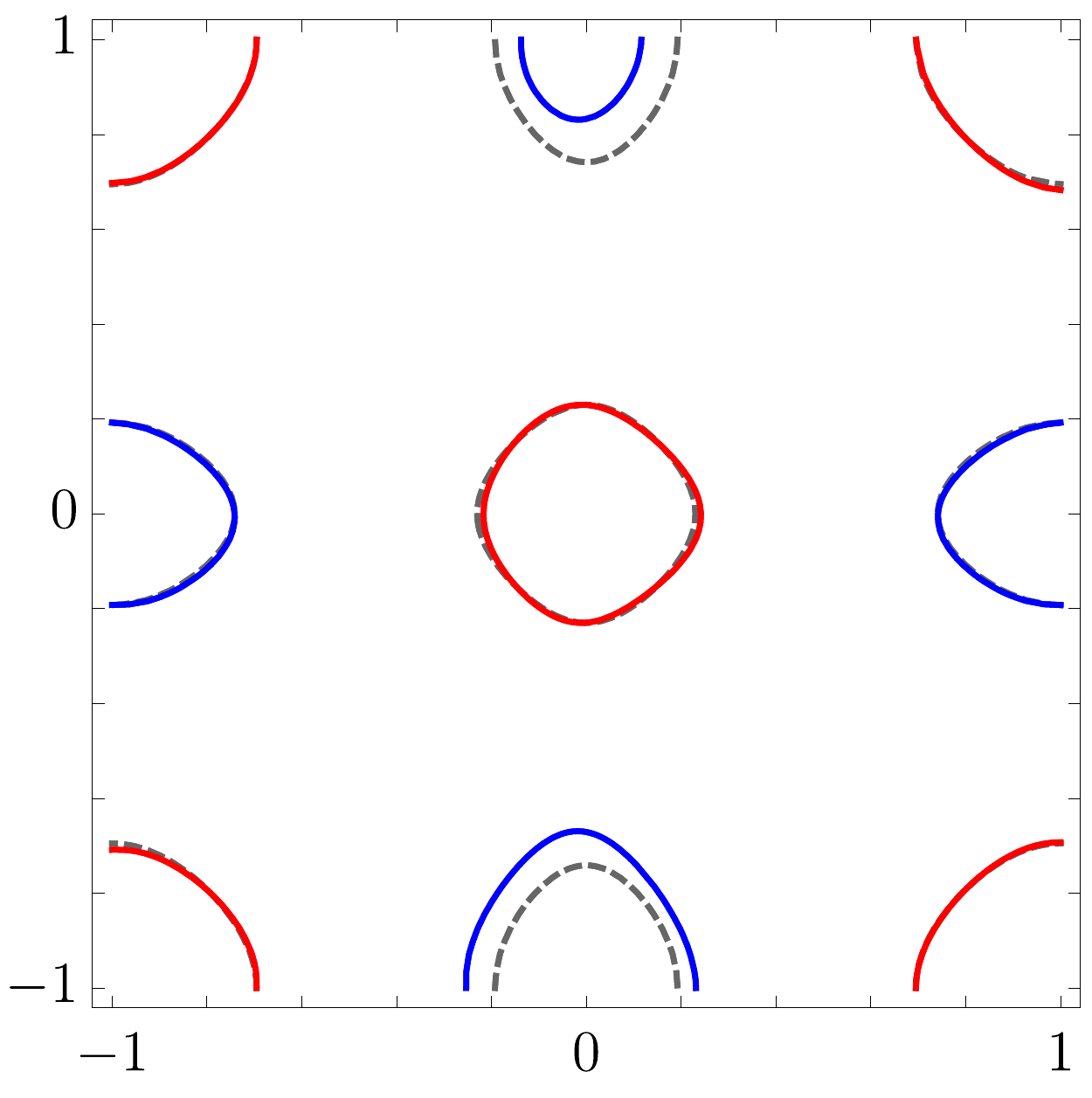}
\end{center}
\end{minipage}\\
\hline
\hline
\end{tabular}\ \ \
\renewcommand{\arraystretch}{1.5}
\begin{tabular}{C{1cm}C{2.8cm}C{2.8cm}}
\hline
\hline &$\mathcal{T}=1$&$\mathcal{T}=-1$\\
\hline
\rule{0pt}{40pt} \begin{minipage}{0\textwidth}\begin{center}
 $\phi^{A_{1},\mathcal{T}}_{s}$
\end{center}
\end{minipage}  &
\begin{minipage}{0.2\textwidth}\begin{center}
\includegraphics[scale=0.18]{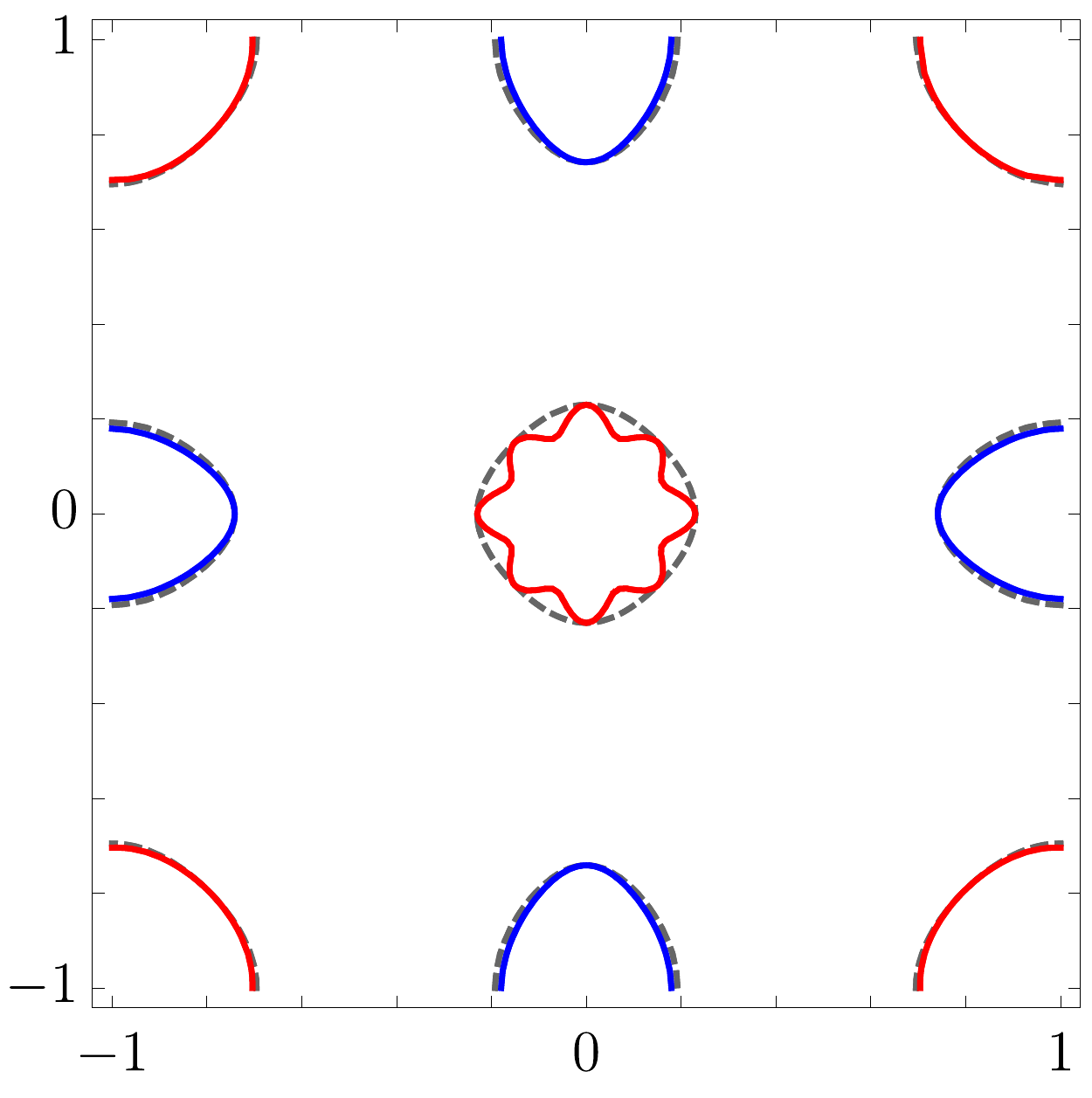}
\end{center}
\end{minipage}
&\begin{minipage}{0.2\textwidth}
\begin{center}
\includegraphics[scale=0.18]{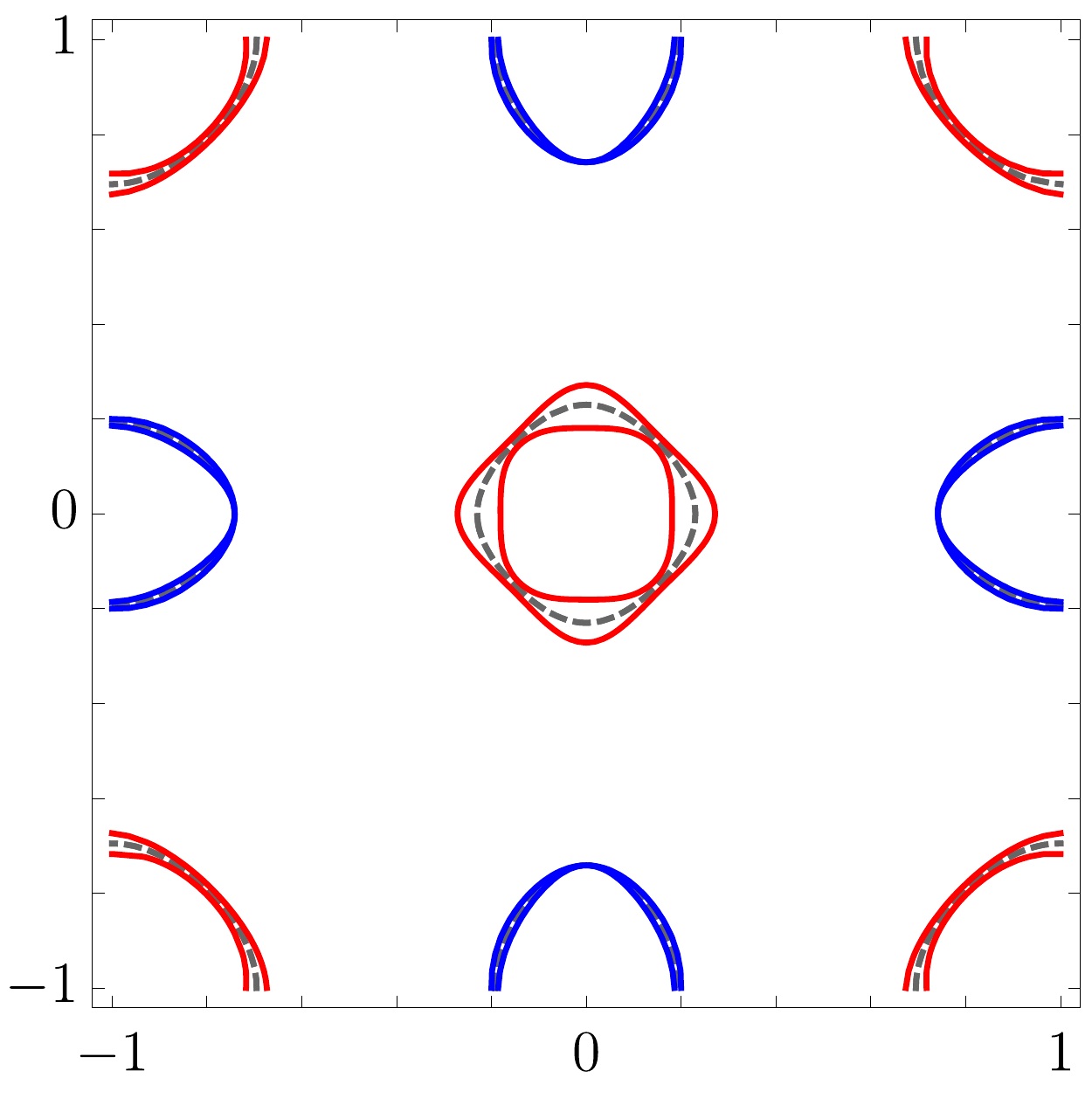}
\end{center}
\end{minipage}
\\
\rule{0pt}{40pt}  \begin{minipage}{0\textwidth}\begin{center}
 $\phi^{A_{2},\mathcal{T}}_{s}$
\end{center}
\end{minipage}  &
\begin{minipage}{0.2\textwidth}
\begin{center}
\includegraphics[scale=0.18]{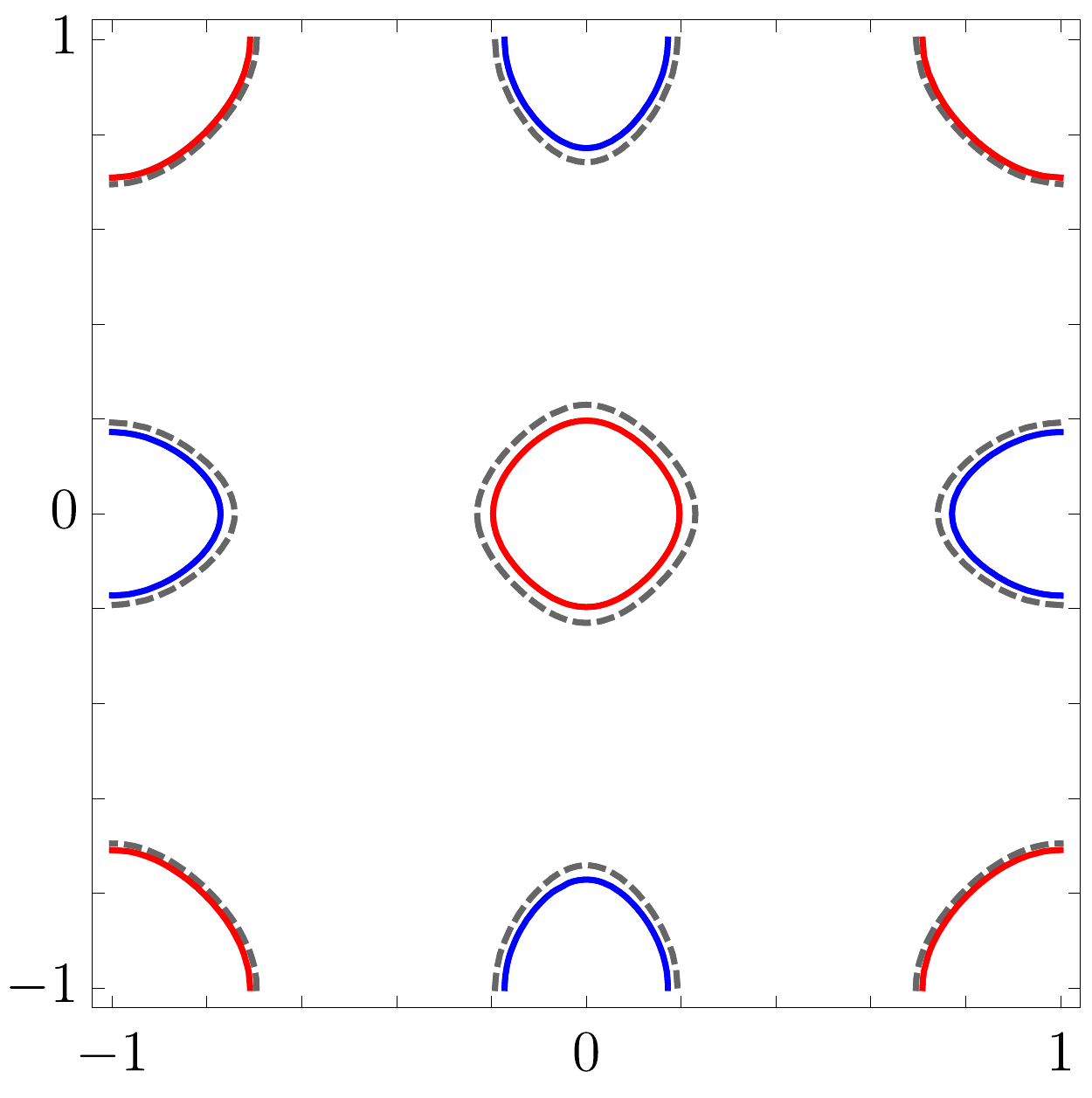}
\end{center}
\end{minipage}
&\begin{minipage}{0.2\textwidth}
\begin{center}
\includegraphics[scale=0.18]{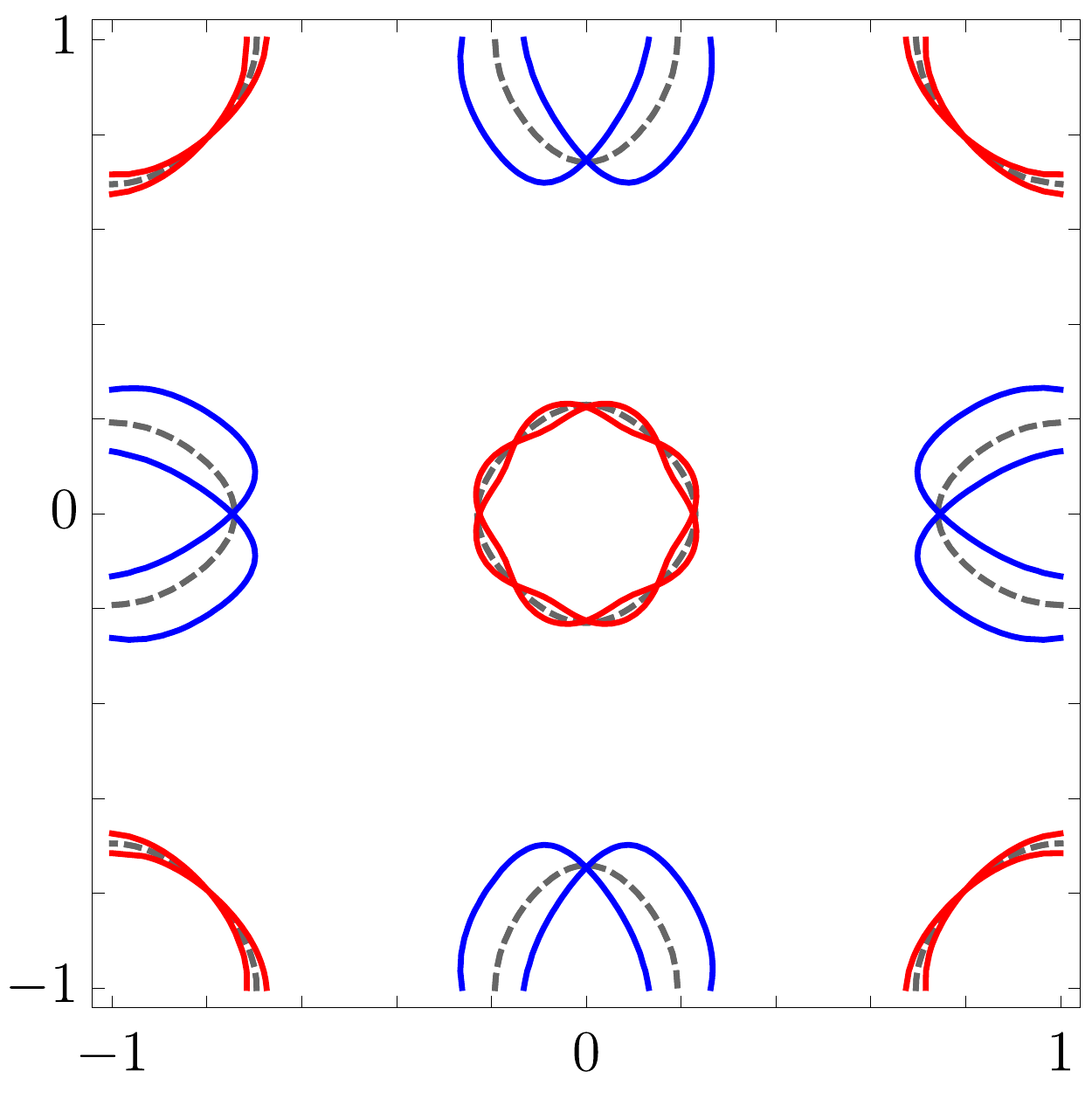}
\end{center}
\end{minipage}\\
\rule{0pt}{40pt}  \begin{minipage}{0\textwidth}\begin{center}
 $\phi^{B_{1},\mathcal{T}}_{s}$
\end{center}
\end{minipage}  &
\begin{minipage}{0.2\textwidth}\begin{center}
\includegraphics[scale=0.18]{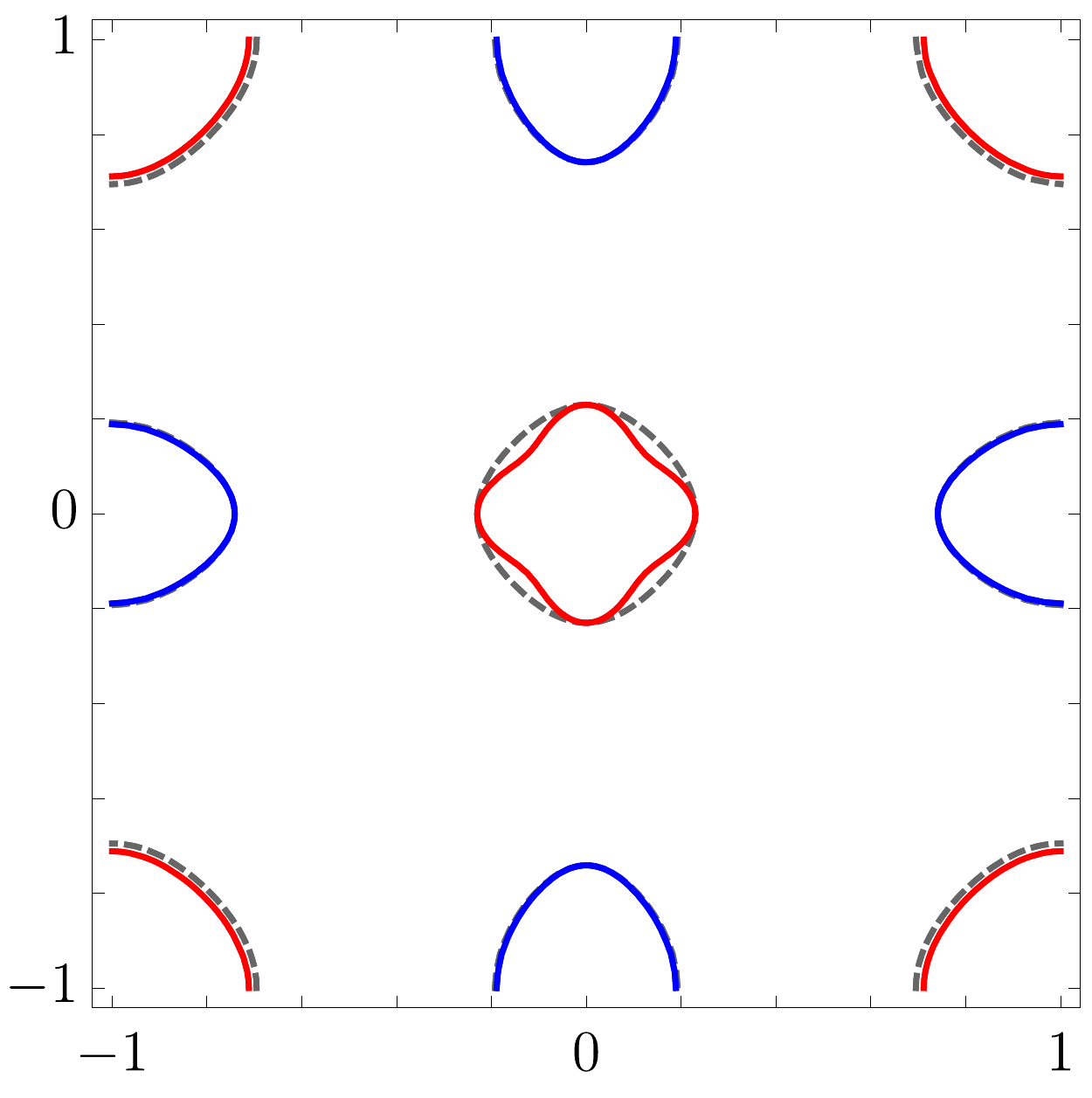}
\end{center}
\end{minipage}
& \begin{minipage}{0.2\textwidth}
\begin{center}
\includegraphics[scale=0.18]{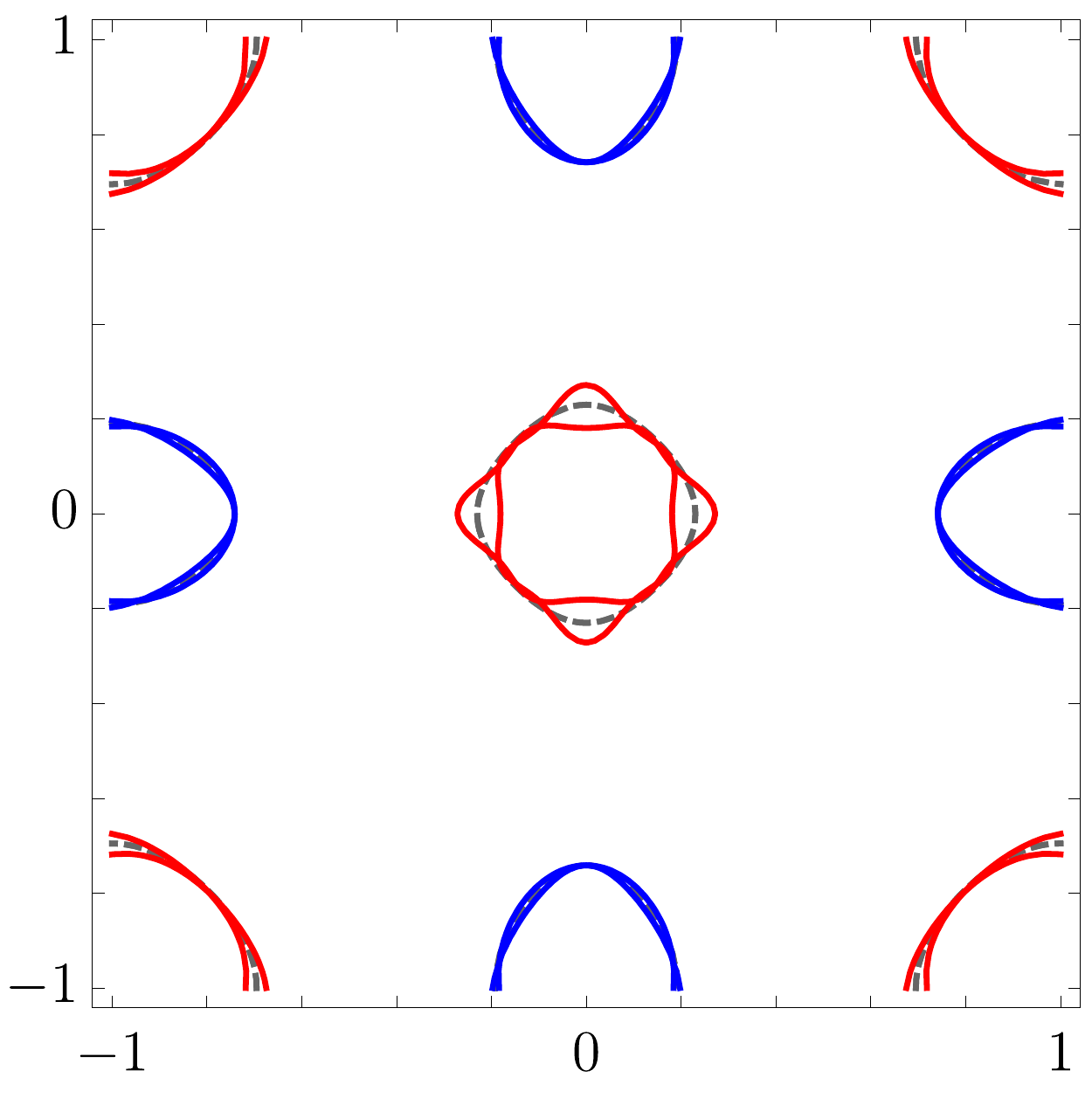}
\end{center}
\end{minipage}\\
\rule{0pt}{40pt}  \begin{minipage}{0\textwidth}\begin{center}
 $\phi^{B_{2},\mathcal{T}}_{s}$
\end{center}
\end{minipage} &
\begin{minipage}{0.2\textwidth}\begin{center}
\includegraphics[scale=0.18]{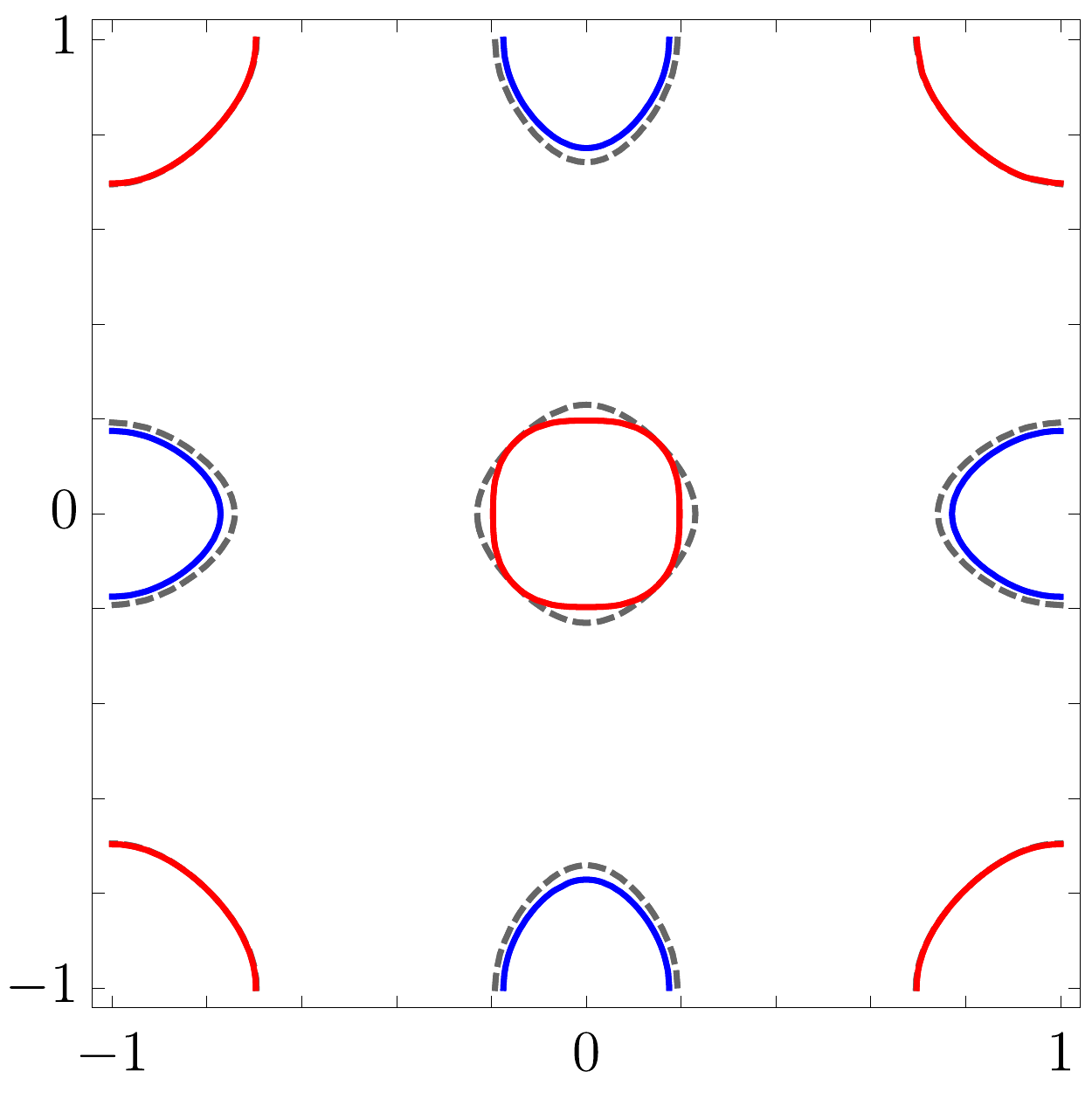}
\end{center}
\end{minipage}
& \begin{minipage}{0.2\textwidth}
\begin{center}
\includegraphics[scale=0.18]{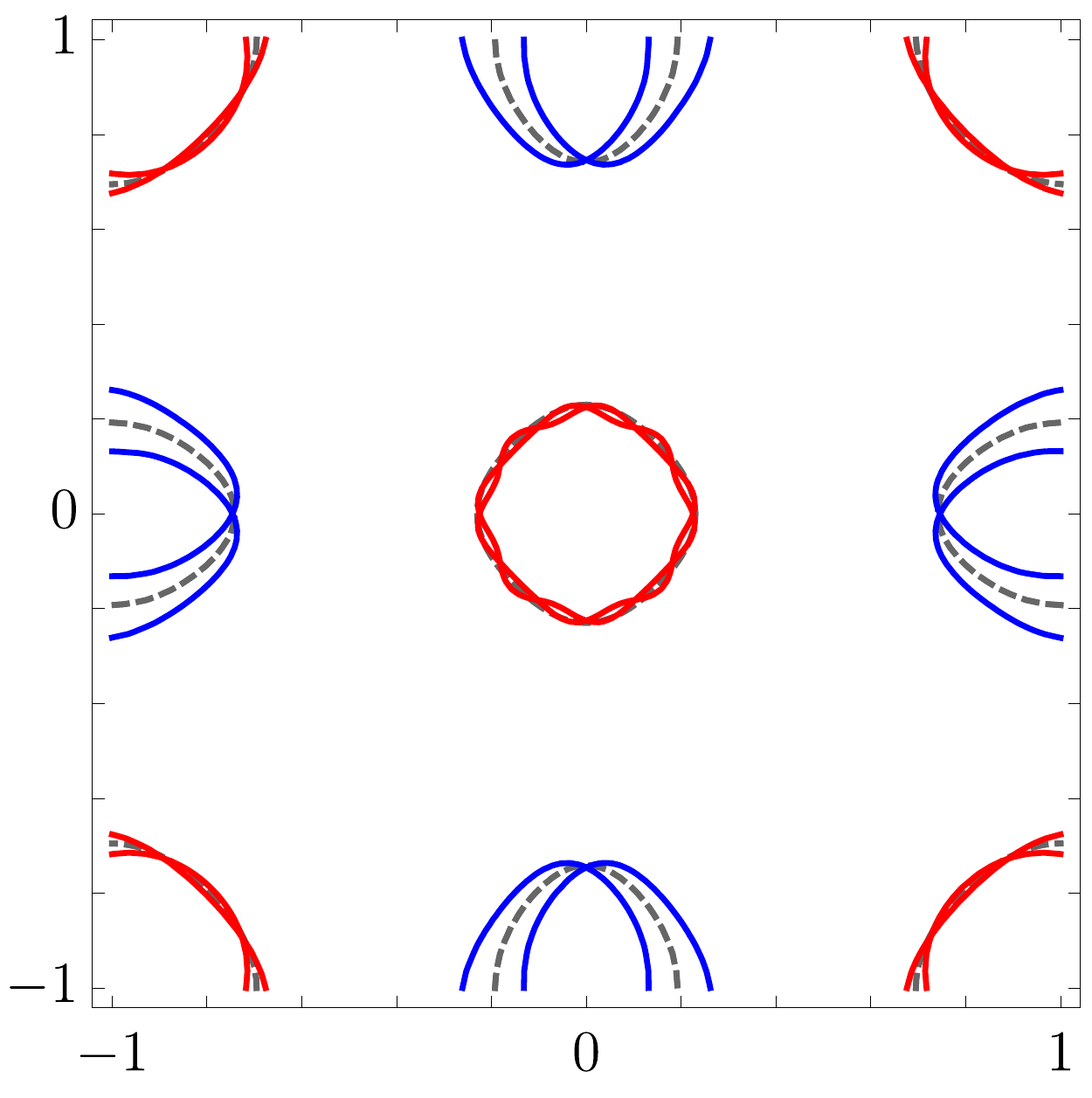}
\end{center}
\end{minipage}\\
\rule{0pt}{40pt} \begin{minipage}{0\textwidth}\begin{center}
 $\phi^{E,\mathcal{T}}_{s}$
\end{center}
\end{minipage}  &
\begin{minipage}{0.2\textwidth}\begin{center}
\includegraphics[scale=0.18]{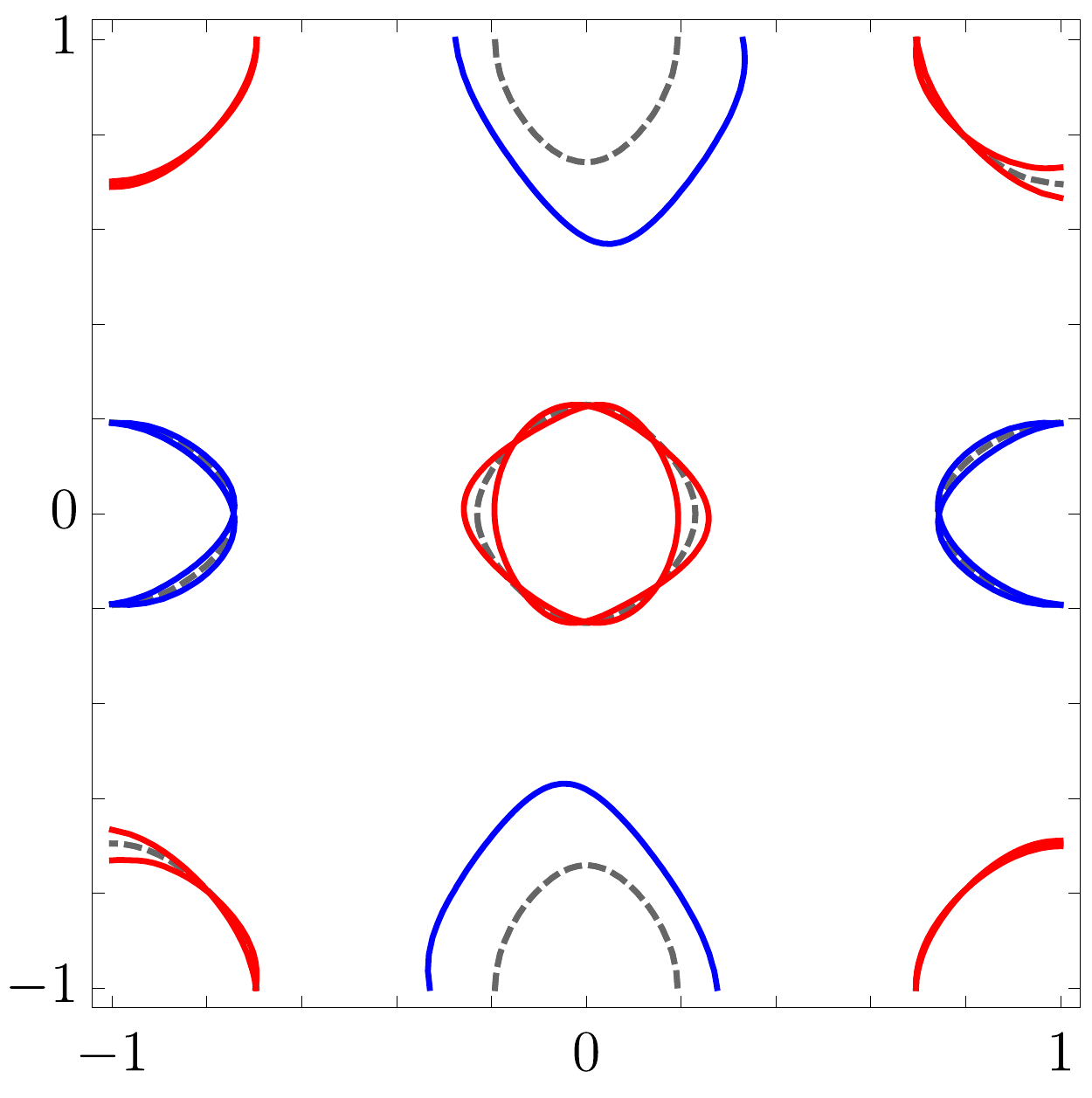}
\end{center}
\end{minipage}
& \begin{minipage}{0.2\textwidth}
\begin{center}
\includegraphics[scale=0.18]{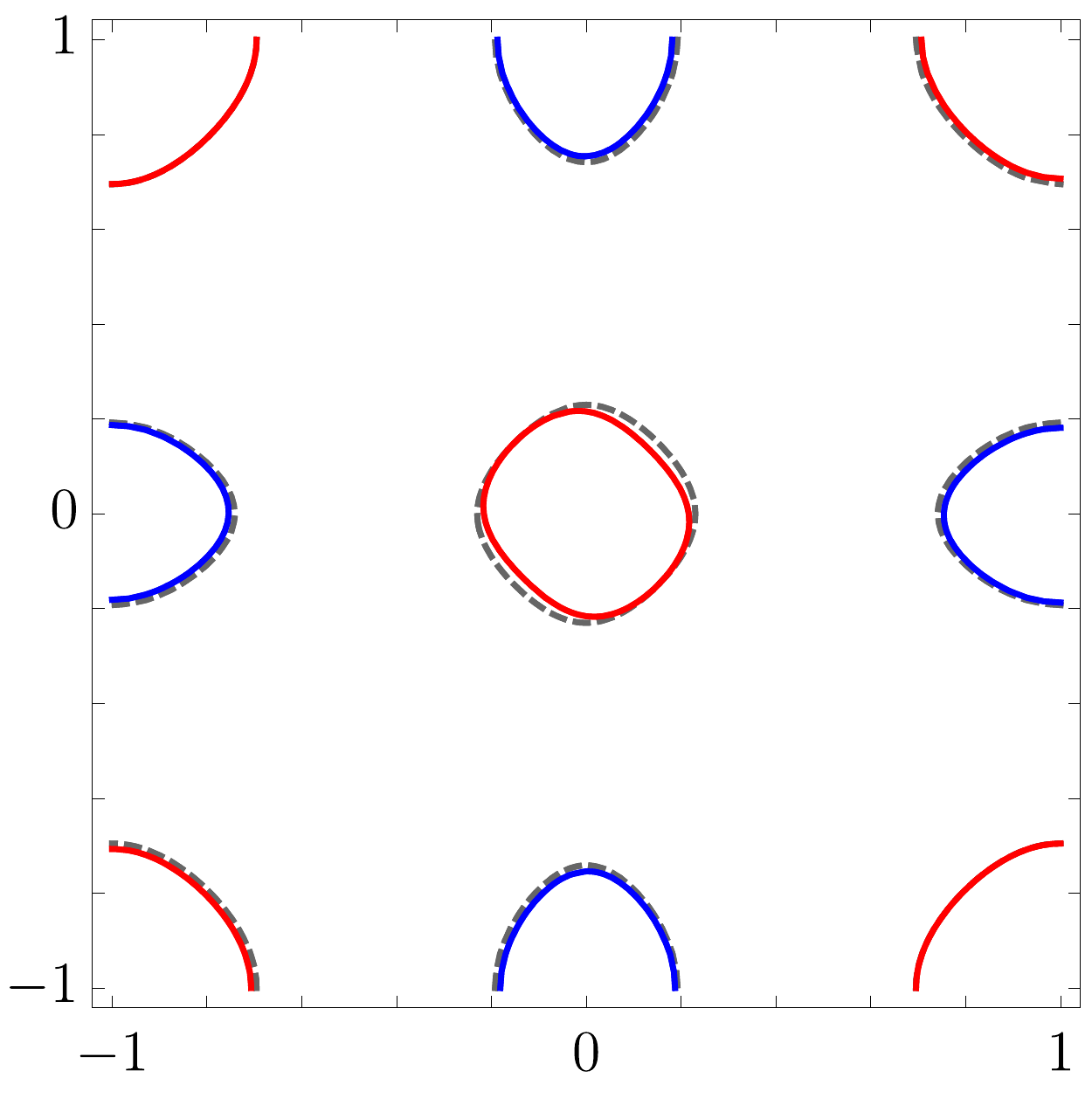}
\end{center}
\end{minipage}\\
\hline
\hline
\end{tabular}

\caption{Illustration of FSs which couple to order parameters, $\phi_{o,s}^{\Gamma,\mathcal{T}}$ in the unfolded BZ. The original FSs (dashed gray line) reconstruct shifted electron (blue line) and hole (red line) pocket. We set a specific choice of a hopping parameter, an order parameter amplitude, and a spin texture.}\label{t3}
\end{center}
\label{tableS1}
\end{table}